\documentclass[fleqn,usenatbib]{mnras}

\usepackage{newtxtext,newtxmath}

\usepackage[T1]{fontenc}

\DeclareRobustCommand{\VAN}[3]{#2}
\let\VANthebibliography\thebibliography
\def\thebibliography{\DeclareRobustCommand{\VAN}[3]{##3}\VANthebibliography}

\usepackage{graphicx}	
\usepackage{amsmath}	

\title[The globular clusters and star formation history of DGSAT~I]{The globular clusters and star formation history of the isolated, quiescent ultra-diffuse galaxy DGSAT~I}

\author[S. R. Janssens et al.]{
Steven~R.~Janssens,$^{1,2}$\thanks{E-mail: sjanssens@swin.edu.au} Aaron~J.~Romanowsky,$^{3,4,5}$ Roberto~Abraham,$^{2}$ Jean~P.~Brodie,$^{1,5}$ Warrick~J.~Couch,$^{1}$
\newauthor{Duncan~A.~Forbes,$^{1}$ Seppo~Laine,$^{6}$ David~Mart\'inez-Delgado$^{7}$ and Pieter~G.~van~Dokkum$^{8}$}
\\
$^{1}$Centre for Astrophysics and Supercomputing, Swinburne University, Hawthorn VIC 3122, Australia\\
$^{2}$Department of Astronomy and Astrophysics, University of Toronto, 50 St. George Street, Toronto, ON, M5S 3H4, Canada\\
$^{3}$Department of Physics and Astronomy, San Jos\'e State University, One Washington Square, San Jose, CA 95192, USA\\
$^{4}$Department of Astronomy and Astrophysics, University of California, Santa Cruz, CA 95064, USA\\
$^{5}$University of California Observatories, 1156 High Street, Santa Cruz, CA 95064, USA\\
$^{6}$IPAC, Mail Code 314-6, Caltech, 1200 E. California Blvd., Pasadena, CA 91125, USA\\
$^{7}$Instituto de Astrof\'isica de Andaluc\'ia, CSIC, Glorieta de la Astronom\'\i a, E-18080, Granada, Spain\\
$^{8}$Department of Astronomy, Yale University, 260 Whitney Avenue, New Haven, CT 06511, USA\\
}

\date{Accepted XXX. Received YYY; in original form ZZZ}

\pubyear{2022}

\begin{document}
\label{firstpage}
\pagerange{\pageref{firstpage}--\pageref{lastpage}}
\maketitle

\begin{abstract}
We investigate the isolated, quiescent ultra-diffuse galaxy (UDG) DGSAT~I and
    its globular cluster (GC) system using two orbits of \textit{Hubble Space
    Telescope} Advanced Camera for Surveys imaging in the
    F606W and F814W filters.
    This is the first study of GCs around a UDG in a low-density environment.
    DGSAT~I was previously found to host an irregular blue low surface
    brightness clump, that we confirm as very likely belonging to the galaxy
    rather than being a chance projection, and represents a recent episode of
    star formation (${\sim}500~\mathrm{Myr}$) that challenges some UDG formation scenarios.
    We select GC candidates based on colours and magnitudes, and construct a
    self consistent model of the GC radial surface density profile along with
    the background.
   We find a half-number radius of
    $R_\mathrm{GC} = 2.7\pm0.1~\mathrm{kpc}$ (more compact than the diffuse starlight)
    and a total of $12 \pm 2$ GCs.
    The total mass fraction in GCs is relatively high,
    supporting an overmassive dark matter halo as also
    implied by the high velocity dispersion previously measured.
    The GCs extend to higher luminosities than expected, and have colours that
    are unusually similar to their host galaxy colour, with a very narrow spread--all of which suggest
    an early, intense burst of cluster formation.
    The nature and origin of this galaxy remain puzzling, but the most likely
    scenario is a ``failed galaxy'' that formed relatively few stars
    for its halo mass,
    and could be related to cluster UDGs whose size and quiescence pre-date their infall.
\end{abstract}


\section{Introduction}

Ultra-diffuse galaxies (UDGs) have attracted considerable attention in recent years, since they were described as an abundant population in the Coma cluster by \citet{vandokkum2015} and were subsequently discovered in many different environments, from clusters to groups to the field \citep[e.g.][]{vanderburg2016,janssens2017,janssens2019,leisman2017,roman2017}.
A variety of mechanisms have been identified that could produce dwarf galaxies 
resembling UDGs, with large sizes (effective radii $R_{\rm e} \gtrsim 1.5$~kpc) and 
low surface brightness (central $\mu_{g,0} \gtrsim $~24~mag~arcsec$^{-2}$).
Conventional explanations include
tidal heating \citep[e.g.][]{yozin2015,martin2019,carleton2019,sales2020},
high spin \citep[e.g.][]{amorisco2016,rong2017},
internal feedback \citep[e.g.][]{dicintio2017,chan2018},
unusual merger histories \citep[e.g.][]{wright2021}
or combinations of these effects
\citep[e.g.][]{jiang2019,liao2019}.
More unconventional is the idea that at least some UDGs are ``failed galaxies''
that stopped forming stars at an unusually early epoch \citep[e.g.][]{beasley2016a,peng2016,vandokkum2019,danieli2022}.

These formation scenarios can be tested through detailed observations of UDGs, going beyond basic properties such as size and luminosity.
Among the most important clues are globular cluster (GC) systems,
which in some UDGs are more populous than in normal dwarfs
(e.g.\ \citealt{beasley2016a,lim2018,danieli2022}).
Recent analysis of the GC systems of six particularly large and luminous UDGs in Coma,
using {\it Hubble Space Telescope} ({\it HST}) imaging,
appeared to rule out several of the more prominent formation scenarios,
leaving failed-galaxies and early-mergers as the best candidates
\citep{saifollahi2022}.

The failed galaxy scenario is promising for explaining a number of otherwise puzzling properties of some UDGs, such as 
their populous GC systems, their
peculiar stellar abundance patterns \citep{ferre-mateu2018,villaume2022},
and their quiescence in low-density environments \citep{papastergis2017}.
The underlying physical picture is that an intense early epoch of star and cluster formation self-quenched these galaxies and left them as quiescent fossil remnants of the early universe.
Some cluster UDGs would then be a product of nature 
rather than nurture, having been formed already as quiescent UDGs before falling in
\citep{vandokkum2019,gannon2022}.

Competing explanations for some of these observations do exist,
such as early infall into clusters to explain populous GC systems
\citep{carleton2021}, and
splashback to explain quenched field UDGs 
\citep{benavides2021}.
Therefore, one powerful discriminant could be observations of GCs around 
quenched field UDGs, to test whether or not these galaxies 
could be the ``missing link'' progenitors of GC-rich UDGs in clusters.

DGSAT~I is the best-studied example of a quenched UDG in a low density environment,
discovered serendipitously as a background galaxy in a study of the M31 halo
\citep{MD2016}.
It is associated with the
Pisces--Perseus supercluster at a distance $D \approx 80~\mathrm{kpc}$,
is red ($V-I =1.0$) and quiescent \citep{MD2016}
and also deficient in cold gas \citep{papastergis2017}.
However, there is an irregular blue overdensity near its centre that may imply a recent episode of star formation \citep{MD2016}.
\citet{MN2019} carried out integral-field spectroscopy of DGSAT~I and
measured a velocity dispersion of 
$\sigma = 56 \pm 10~\mathrm{km}~\mathrm{s}^{-1}$,
which is currently the highest known for any UDG.
They also found a bizarre elemental abundance pattern,
with very low iron content and a very high magnesium-to-iron ratio, that is so far unique among all galaxies.

DGSAT~I is thus an intriguing object with the potential to unlock persistent questions about UDG origins.
Here we present new {\it HST} Advanced Camera for Surveys (ACS)
optical imaging of this galaxy, along with reanalysis of {\it Spitzer} Infrared Array Camera (IRAC)
near-infrared imaging presented in \citet{pandya2018}.
The primary observational goals are
to probe the nature of the central overdensity
and to map out the properties of the GC system--thereby testing whether or not DGSAT~I is an analogue of the
GC-rich UDGs found in higher density environments.

We adopt a distance of $78 \pm 1~\mathrm{Mpc}$ to DGSAT~I
\citep{MD2016}.
This corresponds to a distance modulus of $34.46 \pm 0.03$ and a
physical scale of $0.358~\mathrm{kpc}~\mathrm{arcsec}^{-1}$.
All magnitudes are in the AB system unless otherwise stated.
Galactic extinction corrections from the \cite{schlafly2011} extinction maps
were applied to all colours and magnitudes.\footnote{Using the online
calculator at \url{https://ned.ipac.caltech.edu/forms/calculator.html}.}

This paper is organized as follows.
In Section \ref{sec:data_and_methods}, we discuss the \textit{HST}/ACS and \textit{Spitzer}/IRAC observations of DGSAT~I along with our methods, namely the measurement of morphological parameters for DGSAT~I and the photometry and selection of GC candidates.
Section \ref{sec:results} presents our results: the properties of DGSAT~I, its GC system and the nature of the peculiar blue overdensity.
We interpret these findings in Section \ref{sec:disc} and attempt to create a coherent history of DGSAT~I.
Section~\ref{sec:summ} summarizes our findings.

\section{Data and Methods}\label{sec:data_and_methods}

\subsection{\textit{HST} Observations}

DGSAT~I was observed with \textit{HST}/ACS in two filters (F606W and F814W)
for two orbits (one orbit per filter) on 13 December 2016,
as part of the Cycle 24 program GO-14846 (PI: Romanowsky).
Four exposures were obtained in each filter, using a large+small 4-point dither
pattern to both cover the chip gap and to remove hot pixels and cosmic
rays.
The total exposure times were 2352 s in F606W and 2212 s in F814W.
The calibrated, flat fielded and charge-transfer efficiency (CTE) corrected
individual exposures (\texttt{flc} files) were obtained from MAST and combined
using the \texttt{AstroDrizzle} task from DrizzlePac, weighted with an
inverse-variance weight map (IVM).
\texttt{AstroDrizzle} was run a second time weighted by the exposure time,
but only to obtain exposure time maps.

\begin{figure}
	\includegraphics[width=0.50\textwidth]{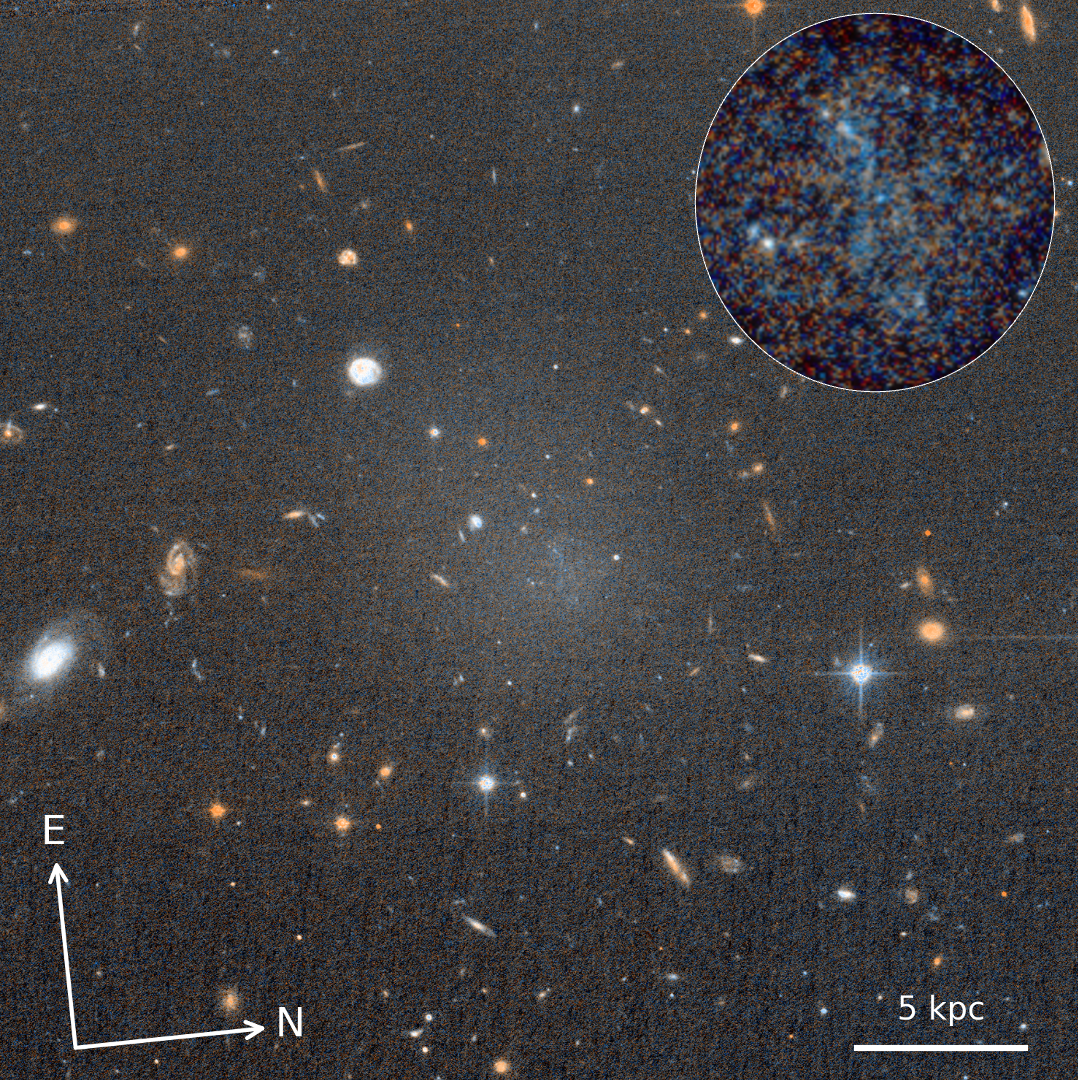}
	\caption{
    \textit{HST}/ACS F606W/F814W colour composite image of DGSAT~I. The scale
    bar in the bottom right is 5 kpc in length, and the image is 32 kpc across
    at the adopted distance to DGSAT~I.
    In addition to the diffuse stellar light, the galaxy hosts a moderately
    rich system of GCs, along with a blue, low surface brightness clump of
    indistinct morphology
    (shown with higher contrast in a zoom-in image at upper right;
    a few GC candidates are also apparent in this zoom-in as fairly ``white'' point sources).
    \label{fig:rgb}
	}
\end{figure}

A portion of the {\it HST} image centred on DGSAT~I is shown in Figure~\ref{fig:rgb}.
Besides the diffuse stellar emission from the galaxy, there are two other features
newly apparent in the {\it HST} imaging.  First is the sprinkling of white
point sources around the central regions, clearly suggesting a population of
GCs that will be analyzed in detail below.  This GC system is not as dramatic
as in similarly large and luminous Coma UDGs such as DF44 and DFX1
\citep{vandokkum2017}, irrespective of how exactly the GC numbers are
quantified.
Second is the central blue clump, offset from the centre of DGSAT~I by ${\sim}5$ arcsec to the northwest (${\sim}1.8~\mathrm{kpc}$ if embedded within the galaxy).
A high contrast zoom-in of the clump is shown as the inset in Figure~\ref{fig:rgb}.
It shows no obvious features
corresponding to a background galaxy, such as spiral arms.
There are, however, hints of very small blue clumps as well as a narrow, jagged
filamentary feature ${\sim}3$~arcsec long (${\sim}1$~kpc if within the galaxy; see Figure~6 of \citealt{MD2016} for the morphology in ground-based imaging).

\subsection{\textit{Spitzer} Observations}

\textit{Spitzer} observed DGSAT~I on 2016 October 23 
as part of the Cycle 16 program 13125 (PI: Romanowsky), in bands IRAC1 and IRAC2--3.6 and 4.5 $\mu$m, respectively--for 1.3 hours in each band.
The \texttt{PBCD} mosaics and uncertainty images for DGSAT~I were obtained from the \textit{Spitzer} Heritage Archive.
These images are in units of $\mathrm{MJy}$ $\mathrm{sr}^{-1}$ which we convert to
$\mu \mathrm{Jy}$~pix$^{-1}$ before analysis.
These observations were previously used in \cite{pandya2018} to analyze the
galaxy's stellar populations, while here we will focus on near-infrared (NIR)
photometry of specific components of the galaxy.

\subsection{DGSAT~I Structural Parameters}\label{sec:galfit}

We used GALFIT \citep{peng2002} to fit a single component S\'{e}rsic model to the 
images of DGSAT~I.
Following \cite{vanderwel2012}, we created a total noise map (``sigma'' image) for GALFIT
by adding the Poisson noise from the image (computed using the
image and the exposure time map) to the rms map (created
from the Drizzle IVM), in quadrature.

We masked other sources for the fit that were
detected in the image by a first pass of \textsc{SExtractor}
\citep{bertin1996} run with default parameters.
This was done by creating a mask image for GALFIT
where pixels within the Kron ellipse of a source (semi-major axis equal to $3
\times \mathtt{A\_IMAGE} \times \mathtt{KRON\_RADIUS}$) were set to 1. Bad
pixels were also masked this way. All other pixels were set to 0.
The masks around bright sources, especially stars and diffraction spikes, were
expanded manually.
While the stacked CTE-corrected images obtained from the archive were sky subtracted
by DrizzlePac, some residual low-order structure remains in the
background.
In particular, there is a valley toward the chip gap that is too complex for
GALFIT to fit with either a fixed sky value or a ramp.
As a result, we restrict GALFIT to the upper third of the image by masking the
rest of the image.
Finally, following \cite{MD2016}, we masked the central blue overdensity
as well.

The galaxy and the background value were then iteratively fit together.
GALFIT was first run assuming a sky background of zero.
The fit axis ratio and position angle were then used to define the shape of the
elliptical annuli used in a curve of growth background value measurement, following the GALAPAGOS
algorithm \citep{barden2012, hiemer2014}.
The curve of growth was computed using the mean pixel value within slightly
overlapping elliptical annuli 60 pixels in width.
A linear slope is computed using the outermost ten measured points.
The sky value is the mean of the second set of ten points that
produces a positive slope, and the standard deviation is adopted as the error.
GALFIT is then re-run using this new sky value, and this process is repeated 5
times, with the fit parameters quickly converging after a few iterations.
We circularized the effective radius using $R_\mathrm{e,c} = R_\mathrm{e} \sqrt{b/a}$, where $R_\mathrm{e}$ is the effective radius along the semi-major axis.
The uncertainties on the model parameters were obtained by refitting DGSAT~I with GALFIT while adopting fixed
background values drawn from the sky value and standard deviation measured above. 

\begin{figure*}
	\includegraphics[width=0.99\textwidth]{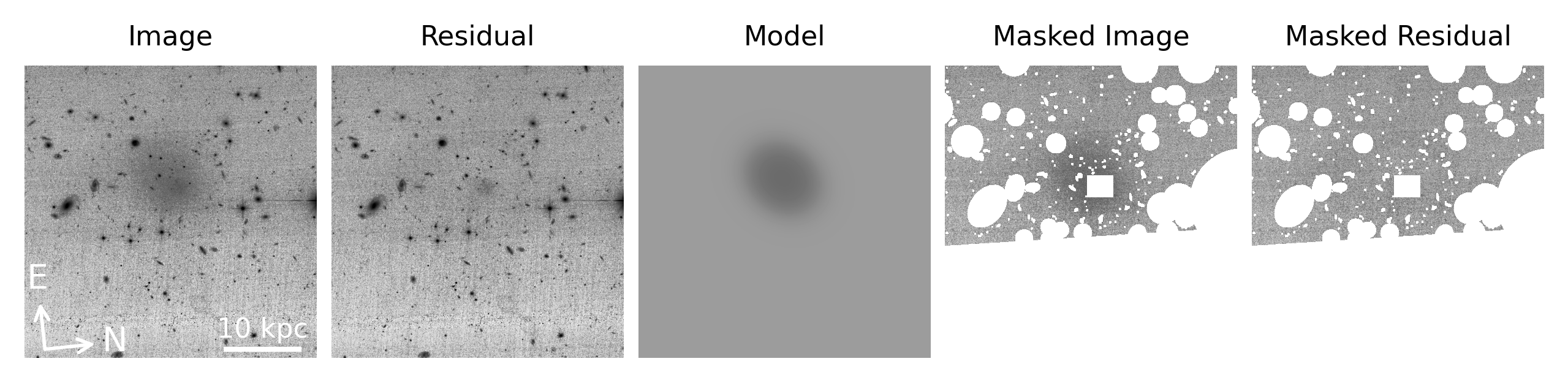}
	\caption{
    Cutouts showing the GALFIT results for the F814W image of DGSAT~I.
    From left to right: original image, residual image, model, masked image and the
    masked residual image.
    The cutouts are 40 kpc $\times$ 40 kpc in size.
    The greyscale is a symmetric logarithmic stretch that is linear around the
    background value to capture both the full dynamic range of the image as well as
    the delicate low surface brightness features of DGSAT~I and the
    background.
    The central irregular overdensity is visible in the residual image.
    Note the discontinuity in the background between the upper and lower
    portions of the cutout, parallel to the ACS chip gap near the bottom of
    the image. In order to prevent this from biasing the background
    measurement, the lower two thirds of the full ACS frame were masked.
    \label{fig:galfit}
	}
\end{figure*}

Figure \ref{fig:galfit} shows the F814W image of DGSAT~I, the
final GALFIT model and residual images, and the masked versions of the original
and residual images.
The central overdensity is most easily seen in the unmasked residual image.
The F606W image was then fitted independently in an identical manner, with
some additional ghosting masked on the right edge of the image.
The $\mathrm{F606W} - \mathrm{F814W}$ colour was measured from the GALFIT model
images using a 1 kpc diameter circular aperture centred on DGSAT~I's fit
position.

GALFIT was also run on the \textit{Spitzer}/IRAC 3.6 and 4.5 $\mu m$ imaging, in order
to examine the NIR properties of the overdensity 
after subtracting a smooth model for the galaxy.
First, \textsc{SExtractor} was run on the \textit{Spitzer} images, and a mask
image was created in an identical manner to the \textit{HST} imaging discussed
above (though here the semi-major axis of the Kron ellipse is equal to $1
\times \mathtt{A\_IMAGE} \times \mathtt{KRON\_RADIUS}$).
Likewise, the mask for the blue overdensity was transformed to the
\textit{Spitzer} image and applied as well.
Cutouts $189{\arcsec} \times 189{\arcsec}$ in size centred on DGSAT~I were
made from the mosaic, uncertainty and mask images.
The uncertainty cutout was passed to GALFIT as the ``sigma'' image.

We created a point spread function (PSF) model for each IRAC band using the standalone version of DAOPHOT
\citep{stetson1987}.
First, the mosaic image was transformed into units of electrons using the
\texttt{FLUXCONV}, \texttt{GAIN} and \texttt{EXPTIME} values in the header.
The PSF was then modeled by DAOPHOT by fitting a Gaussian function to
well-exposed isolated stars.
No spatial variation of the PSF across the image was adopted.
A ``fitting radius'' of 5 pixels was used, along with a ``PSF radius'' of 51
pixels.
The \texttt{ADDSTAR} task was then used to create a PSF image for GALFIT.

A single component S\'{e}rsic model was then fit to DGSAT~I with all
parameters but the centroid, position angle and magnitude fixed to the
\textit{HST} F814W values.
In addition, all neighbours brighter than $m_\mathrm{IRAC1} =
22~\mathrm{mag}$ within $50{\arcsec}$ of DGSAT~I were simultaneously fitted with PSF models (though
several neighbours are background galaxies in the \textit{HST} imaging, they are all
unresolved by \textit{Spitzer}).
The overdensity is examined in the GALFIT residual images and this is
discussed later in Section \ref{sec:overdensity}.

\subsection{Photometry, Artificial Star Tests and Point Source Selection}
\label{sec:phot}

We ran \textsc{SExtractor} in dual-image mode on the F606W and F814W
CTE-corrected images obtained from MAST, with the F814W image used as the
detection image.
Source magnitudes were measured in apertures 4 pixels in diameter,
with the background measured in local annuli using default background parameters.

The effective PSF prescription described in
\cite{anderson2000}, and implemented in Photutils \citep{bradley2021}, was used to create a model
PSF for the F606W and F814W images from 32 isolated, bright, but unsaturated
stars in each image.
Aperture corrections were then applied by first correcting to an aperture of
{1\arcsec} diameter using the model PSF, and then to infinity using Table 5 in
\cite{sirianni2005}.
The total corrections were 0.55 and 0.71 mag in F606W and F814W, 
respectively.\footnote{Alternatively, using a PSF constructed with DAOPHOT would yield corrections of 0.63 and 0.75 mag. These differences imply that the absolute colour values have a systematic uncertainty at the $\sim$~0.05 mag level.}

\begin{figure}
	\includegraphics[width=0.50\textwidth]{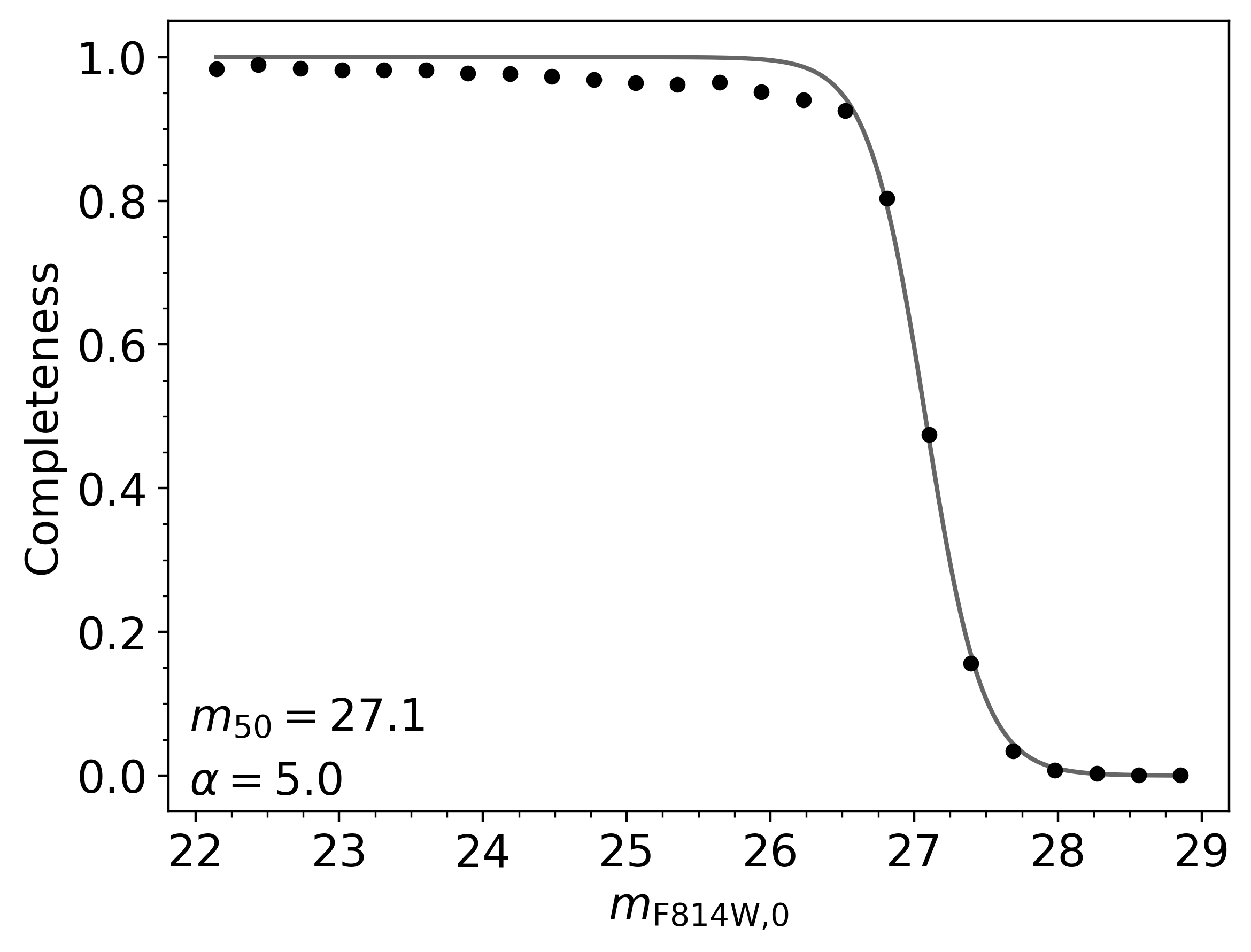}
	\caption{
    Completeness as a function of F814W magnitude.
    Points show results for artificial stars in bins
    0.3 mag in size.
    The solid line is the best fit function following Equation~\ref{eqn:fm}.
    We find a 50\% completeness limit in F814W of 27.1 mag.
    \label{fig:completeness}
	}
\end{figure}

We then injected 50,000 artificial stars with magnitudes $22 < m_\mathrm{F814W}
< 29$.
onto blank areas of the
F814W images in batches of 2000.
Figure \ref{fig:completeness} shows the recovered fraction of artificial stars
in bins of F814W magnitude.
The completeness is modeled using the function
\begin{equation}\label{eqn:fm}
f(m) = \frac{1}{1 + e^{\alpha(m - m_{50})}},
\end{equation}
where $m_{50}$ is the magnitude at which the completeness falls to 50\% and
$\alpha$ controls how steeply the completeness drops off \citep{harris2016}.
The function deviates slightly from the data at the bright end because of losses to blending.
We find a 50\% completeness limit of $m_{\mathrm{F814W},0} = 27.1$.

\begin{figure}
	\includegraphics[width=0.50\textwidth]{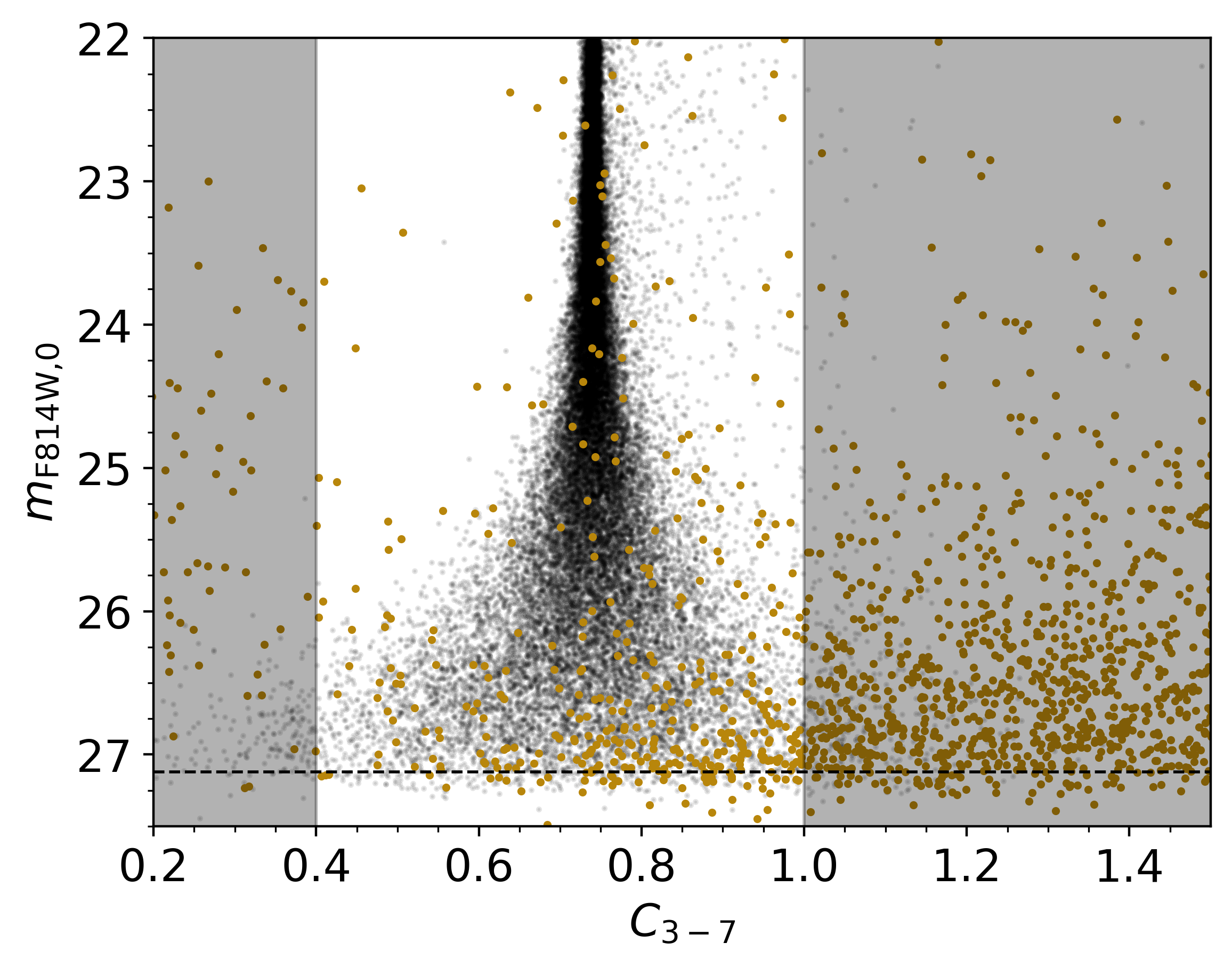}
	\caption{
    Point source selection using image concentration, $C_{3-7}$, defined
    as the difference between an object's magnitude measured in a 3 and 7
    pixel diameter aperture. Shown in black are artificial stars injected into
    the image, while real sources are shown in gold. The horizontal dashed black line is the
    50\% completeness limit.
    The unshaded region is our adopted point source selection of $0.4 <
    C_{3-7} < 1.0$.
    \label{fig:concentration}
	}
\end{figure}

Finally, the artificial stars are used to define our point source selection.
Figure \ref{fig:concentration} shows F814W magnitude plotted against image
concentration, $C_{3-7}$, defined as the difference between an object's
magnitude measured in 3 and 7 pixel diameter apertures.
At the adopted distance to DGSAT~I, these apertures correspond to 54 pc and
125 pc, respectively.
The black points are our artificial stars while real detections in the image
are plotted in gold.  We adopt a conservative point source selection of $0.4 <
C_{3-7} < 1.0$.

\subsection{Globular Cluster Selection}

\begin{figure*}
	\includegraphics[width=0.99\textwidth]{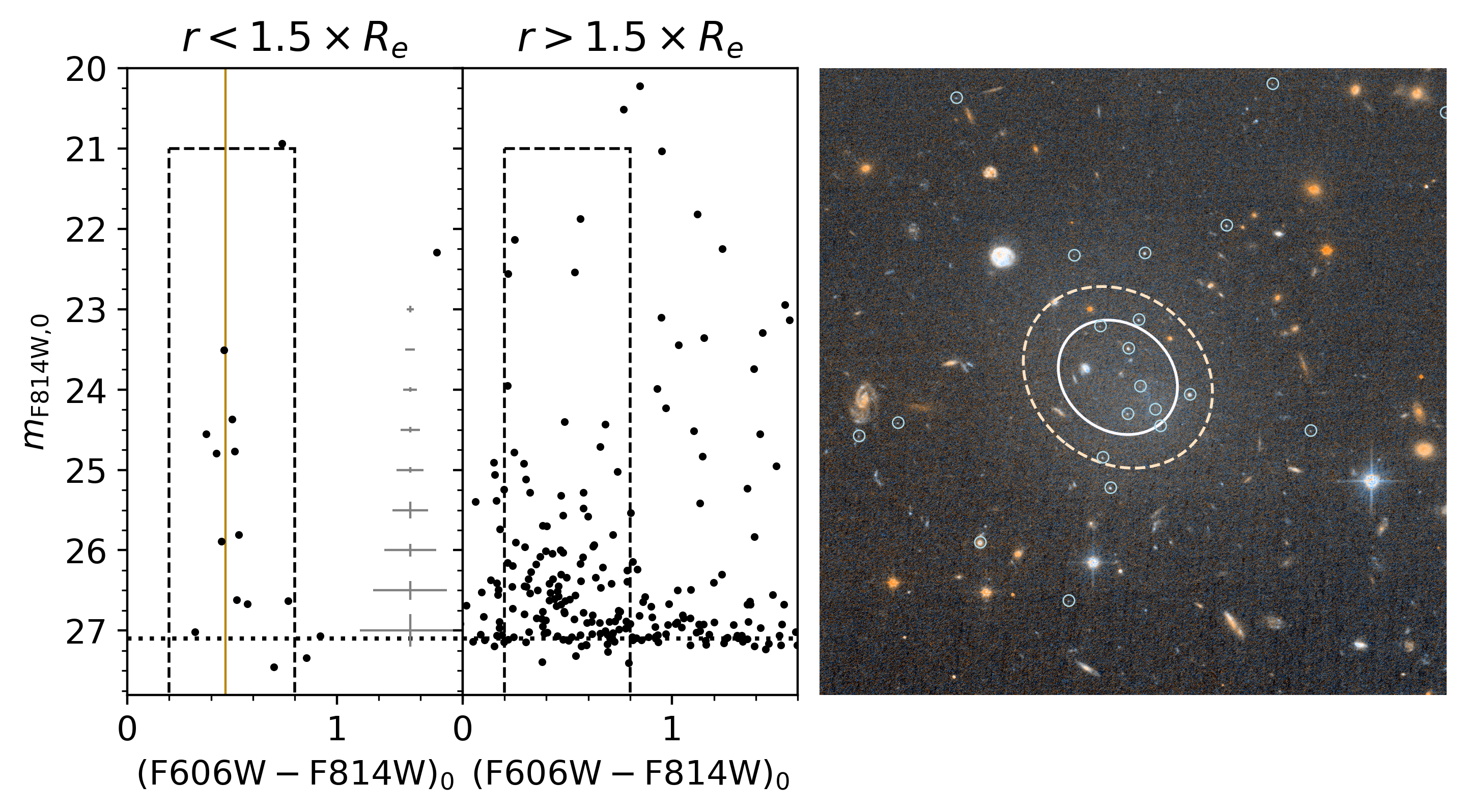}
	\caption{
    \textit{Left}: F606W/F814W colour--magnitude diagrams (CMDs) of point
    sources in proximity to DGSAT~I (left CMD, $r < 1.5 \times
    R_\mathrm{e}$) and those in the rest of the image (right CMD), covering a ${\sim}125$ times larger area.
    The grey error bars are representative uncertainties in 0.5 magnitude bins from
    the artificial star tests.
    Point sources near DGSAT~I predominantly occupy a narrow range in both
    magnitude and colour, as expected for GCs.
    We adopt a colour cut of $0.2 < \mathrm{F606W} - \mathrm{F814W} < 0.8$ 
    and a bright magnitude limit of $\mathrm{F814W} = 21$ to select GC candidates.
    This GC selection box is shown as the dashed rectangle.
    The horizontal dotted line is the 50\% completeness limit of $\mathrm{F814W} = 27.1$.
    The vertical solid gold line marks our measured colour for the diffuse stellar light
    of DGSAT~I ($\mathrm{F606W} - \mathrm{F814W} = 0.46$),
    which is remarkably similar to the GC colours.
    A bright UCD candidate is at $\mathrm{F814W} \sim 21$, but it is unresolved so we believe it to be a foreground star (see text for details).
    \textit{Right}: F606W/F814W colour composite zoom-in 
    on DGSAT~I with GC
    candidates marked with small, light blue circles.
    The cutout is $25~\mathrm{kpc} \times 25~\mathrm{kpc}$ in size.
    The half-light boundary
    $R_\mathrm{e}$ of the galaxy is outlined with the outer dashed ellipse, and
    the half-number boundary of the GC system is outlined with the inner solid
    ellipse.
    The UCD candidate is located on the outer ellipse at upper left. 
    \label{fig:cmd}
	}
\end{figure*}

GC candidates were selected from the point source catalog on the basis of
magnitude and colour.
This is shown on the left side of Figure \ref{fig:cmd}, where we plot
colour--magnitude diagrams (CMDs) for point sources both in the vicinity of DGSAT~I
(distance $r < 1.5 \times R_\mathrm{e}$) and in the rest of the image,
including both chips.
Characteristic error bars derived from the artificial star tests are shown as
the light grey points.

Compared to the rest of the image, point sources in the vicinity of DGSAT~I
occupy a narrow range in colour that is consistent with expectations for GCs.
To define an allowed colour range for GCs, we generated 
simple stellar populations (SSPs) using Flexible Stellar Population Synthesis
(FSPS; \citealt{conroy2009,conroy2010}).
We allow for a relatively wide range of metallicity and age, from
$[\mathrm{Fe}/\mathrm{H}] = -2.0$ to $+0.0$ dex and 5--13~Gyr.
The corresponding colour range is $0.27 < \mathrm{F606W} - \mathrm{F814W} < 0.69$.
We adopt a conservative GC colour cut of $0.2 < \mathrm{F606W} -
\mathrm{F814W} < 0.8$, which is slightly wider to allow for photometric errors, and a
bright magnitude limit of $\mathrm{F814W} = 21$, roughly $M_V \approx -13$, to
select GC candidates.
This GC selection box is shown as the dashed rectangle
in the left panels of Figure~\ref{fig:cmd}.
Point sources satisfying this selection are circled in light blue in the right panel of Figure \ref{fig:cmd}.
We will discuss this Figure further in Section \ref{sec:GCs}.

We also attempt to characterize the sizes of the GCs, which is possible even at $\sim$~75~Mpc distances using careful PSF-convolved modeling packages such as ISHAPE \citep{larsen1999,harris2020}. 
We fit circular \cite{king1962} models with concentration $c \equiv r_t/r_c = 30$ to each of the GC candidates, in the F814W image.
For this measurement alone, the \textit{HST}/ACS imaging was reprocessed with \textsc{AstroDrizzle} to a 0.03 arcsec pixel scale.
ISHAPE also requires a PSF model subsampled by a factor of 10, which we generated using the IRAF-bundled version of DAOPHOT.
The three brightest GCs (with $M_V \sim -10$) are clearly resolved and have 
half-light radii $r_{\rm h} \sim 10~\mathrm{pc}$.
These large sizes contrast with $r_{\rm h} \sim 3~\mathrm{pc}$ for classical GCs, and the objects could arguably be classified as UCDs \citep[e.g.][]{brodie2011,norris2014}.
The sizes of the fainter GCs around DGSAT~I are poorly constrained with the present data, although there are hints that some may also be unusually large.

Most of the GC candidates around DGSAT~I are clustered in colour and magnitude, as will be explored in detail later.
The exception is a relatively bright object at $\mathrm{F814W} \sim 21$, which
is distinct in colour and magnitude from the rest of the GCs.
This object might be a member of the general population of ultra-compact
dwarfs (UCDs), which in many cases deviate from the properties shared by
classical GCs (e.g.\ \citealt{brodie2011}), and have been identified in
association with cluster UDGs \citep{janssens2019}.
The size measurement from ISHAPE however reveals that this object is unresolved.
All compact stellar systems yet known that are this luminous have half-light radii $r_\mathrm{h} \gtrsim 10~\mathrm{pc}$ \citep{sandoval2015}.
We conclude that this object is likely a foreground star and omit it from our GC sample.

A special consideration for contamination in the DGSAT~I observations is this galaxy's
location behind the halo of M31, with the satellite And~II about 15\arcmin\ away.
Foreground red giant branch and horizontal branch stars could in principle
cross through our GC selection box
(see \citealt{mcconnachie2007}).
However, we see no distinct trend signifying
these populations in the middle panel of Figure~\ref{fig:cmd}, where the contaminants
have instead a fairly uniform colour distribution.
We also find no sign of a gradient in contaminants across the {\it HST} field 
that would imply a significant contribution from the declining outer stellar envelope of And~II.

\section{Results}\label{sec:results}

Here we present the results on the structural parameters of the galaxy DGSAT~I
(Section~\ref{sec:galfitresults}),
on the central overdensity (Section~\ref{sec:overdensity})
and on the GC system (Section~\ref{sec:GCs})

\subsection{DGSAT~I Structural Parameters}
\label{sec:galfitresults}

\begin{table}
\caption{DGSAT~I Structural Parameters}
\label{tab:params}
\begin{tabular}{rl}
\hline
$m_\mathrm{F814W}$              & $18.00 \pm 0.06$ \\
$M_\mathrm{F814W}$              & $-16.46 \pm 0.09$ \\
$R_\mathrm{e}$                  & $4.00 \pm 0.08~\mathrm{kpc}$ \\
$R_\mathrm{e,c}$                & $3.67 \pm 0.07~\mathrm{kpc}$ \\
$\mu_{0,\mathrm{F814W}}$        & $24.8 \pm 0.1~\mathrm{mag}~\mathrm{arcsec}^{-2}$ \\
$\langle\mu\rangle_{\mathrm{e},\mathrm{F814W}}$ & $25.0 \pm 0.1~\mathrm{mag}~\mathrm{arcsec}^{-2}$ \\
$n$                             & $0.38 \pm 0.02$ \\
$\mathrm{F606W}-\mathrm{F814W}$ & $0.46 \pm 0.01$ \\
$b/a$                           & $0.840 \pm 0.002$ \\
$\theta$                        & $51.7 \pm 0.6$ deg \\
\hline
\end{tabular}
\end{table}

Table \ref{tab:params} summarizes the results from GALFIT modeling of the galaxy light,
based on the F814W image (Section~\ref{sec:galfit}).
The results from F606W were very similar,
and the implied $V$-band absolute magnitude is $M_V = -15.9$.
Note that these measurements do not include the small flux contributions from
the blue overdensity, which was masked out in the fits.
The luminosity uncertainties include the distance uncertainty, but the sizes do not.
We also tried an alternative method of measuring the galaxy colour
by fixing the GALFIT parameters to be the same in the two bands,
other than the overall flux normalizations
(cf.\ \citealt{cohen2018}).
The result was very close to our aperture colour measurement from
independent fits in the two filters.
There is in principle a $K$-correction of ${\sim}0.01$ mag to the measured colour, given the galaxy's redshift\footnote{\url{http://kcor.sai.msu.ru/}}, but for the remainder of the paper, we do not apply this correction, since it is comparable to the measurement errors.

The values from {\it HST} in Table~\ref{tab:params} are in broad agreement with 
previous ground-based values (where the central overdensity was masked out, as we have also done).
For example, \citet{MD2016} found $R_\mathrm{e} = 4.7 \pm 0.2$~kpc,
$n = 0.6 \pm 0.1$, $b/a = 0.87 \pm 0.01$,
$I = 17.17 \pm 0.05$ and $V-I = 1.0 \pm 0.1$ (Vega),
which are equivalent to 
$\mathrm{F814W} \simeq 17.8$ and $\mathrm{F606W} - \mathrm{F814W} \simeq 0.4$ (AB).
\citet{pandya2018} found $R_\mathrm{e} = 5.1$~kpc and $I = 17.70$ (AB),
which is equivalent to $\mathrm{F814W} \simeq 18.0$. 
They also found $V-I = 0.32 \pm 0.11$ (AB) from aperture photometry,
which translates to F606W$-$F814W = $0.22 \pm 0.09$.
Our measured central surface brightness in F814W is slightly fainter than the $I$-band value of $\mu_{0,I} = 24.0 \pm 0.2~\mathrm{mag}~\mathrm{arcsec}^{-2}$ found by \cite{MD2016}, though this likely stems from the differing S\'{e}rsic indices.

The stellar mass for DGSAT~I was previously estimated based on broad-band colors to be
$M_\star \simeq (4$--$5)\times10^8 M_\odot$,
with a mass-to-light ratio of $M_\star/L_I =$~1.1--1.3 in solar units
\citep{MD2016,pandya2018}. 
Here we use FSPS to estimate $M/L = 1.33$ and $1.63$ in F814W and F606W, respectively,
assuming a \cite{chabrier2003} IMF, 
and using stellar population parameters that will be discussed later.
This gives an estimated $M_\star \simeq 3.3\times10^8 M_\odot$.

\subsection{The Central Overdensity} \label{sec:overdensity}

\begin{figure*}
    \includegraphics[width=0.99\textwidth]{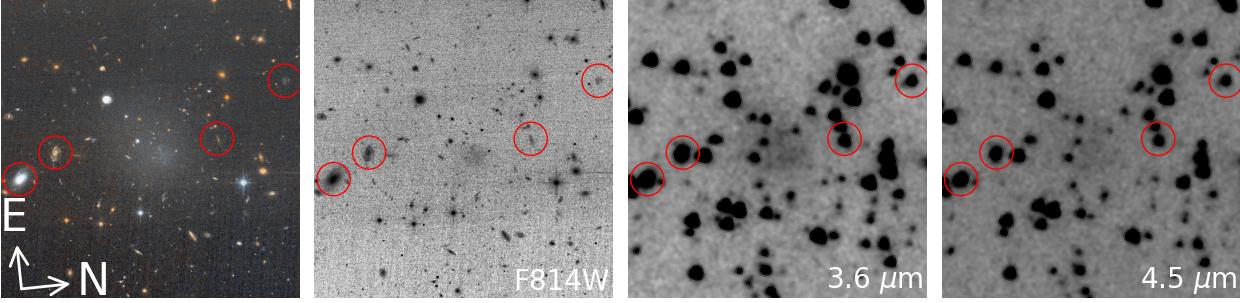}
	\caption{
    The near-infrared appearance of DGSAT~I's overdensity. From left to right
    are shown the F606W/F814W colour composite image, the \textit{HST} F814W
    GALFIT residual image,
    highlighting the overdensity and its position, and the
    \textit{Spitzer} $3.6~\mathrm{\mu m}$ and $4.5~\mathrm{\mu m}$ images.
    The cutouts are 32 kpc $\times$ 32 kpc in size.
    The background galaxies visible in the \textit{HST} imaging all have
    bright counterparts in both \textit{Spitzer} channels.
    To aid visual comparison, a handful of background galaxies with different apparent redshifts and morphologies are marked.
    The faint appearance of the overdensity in the \textit{Spitzer} images
    leads us to conclude it is not a superimposed background galaxy.
    \label{fig:overdensity_spitzer}
	}
\end{figure*}

The \textit{HST} imaging enables a closer look at the slightly offset 
central overdensity of DGSAT~I.
We measured the optical magnitude and colour of this feature on the GALFIT residual images
(since the overdensity was masked when GALFIT was run) in a
$7^{\prime\prime}$ diameter circular aperture centered on the overdensity.
We find $m_\mathrm{F606W} = 21.7$ and $\mathrm{F606W}-\mathrm{F814W} = 0.21$, 
notably bluer than the main body
of the UDG ($\mathrm{F606W}-\mathrm{F814W} = 0.46$; Section~\ref{sec:galfitresults}).
Transforming the overdensity 
colour using the
\cite{saha2011} transformations, we find $V - I = 0.72$, in
agreement with $V-I=0.7$ measured with Subaru/Suprime-Cam \citep{MD2016}.
We do not believe the overdensity to be a background galaxy as
we find little to no counterpart of the overdensity in the \textit{Spitzer} $3.6$ and
$4.5~\mathrm{\mu m}$ imaging of DGSAT~I; all other background galaxies visible
in the \textit{HST} imaging have bright counterparts in the \textit{Spitzer}
images as shown in Figure \ref{fig:overdensity_spitzer}.

\begin{figure*}
    \includegraphics[width=0.99\textwidth]{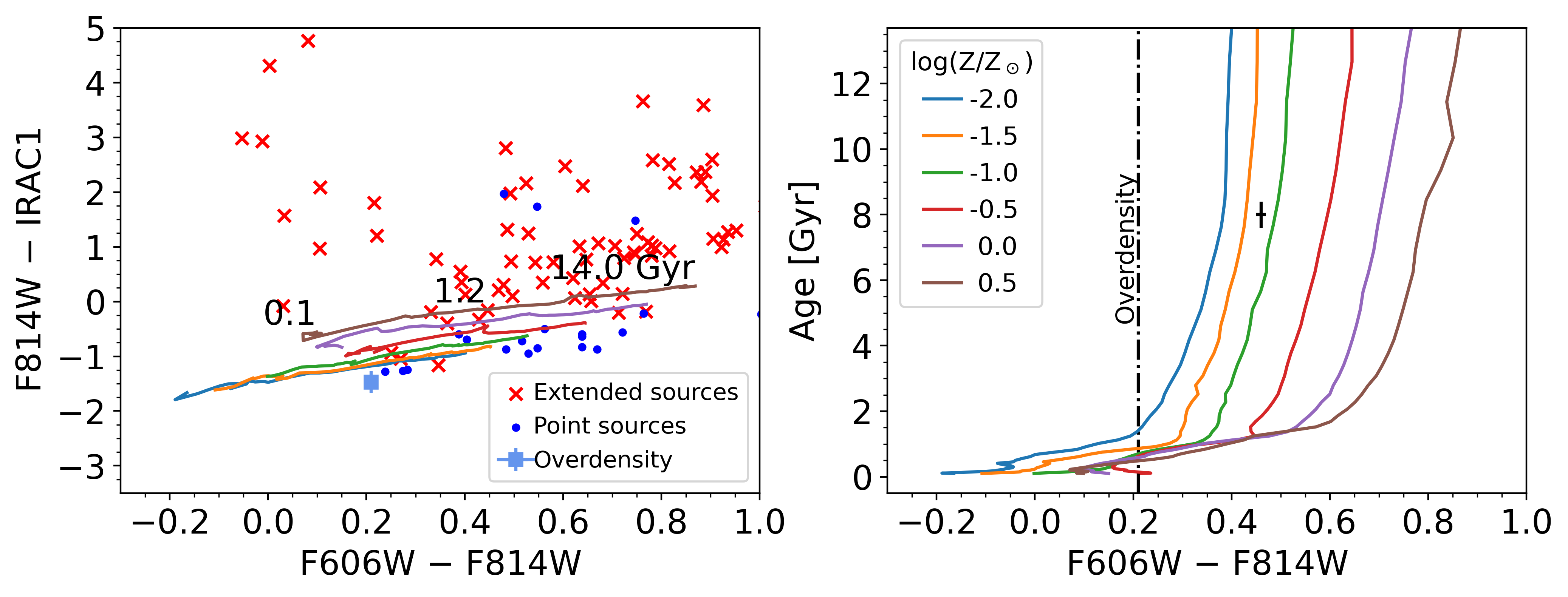}
	\caption{
    \textit{Left}: The colour of the overdensity compared to other sources in the \textit{Spitzer} imaging.
    The blue square is the overdensity.,
    Sources are separated into extended (red crosses, presumed background galaxies) 
    and point sources (blue points), determined from the \textit{HST} imaging.
    The $\mathrm{F814W}-\mathrm{IRAC1}$ colour of the overdensity is bluer than most other galaxies.
    Also shown are SSP tracks of various metallicities from FSPS
    (see legend in right panel),
    labelled in age from left to right.
    \textit{Right}: 
    Age vs.\ optical colour, with the same SSP tracks shown again.
    The black error bars show the colour and spectroscopic age of the main body of DGSAT~I.
    The dash-dotted vertical line is the measured colour of the overdensity,
    which must be young in order to appear so blue.
    The most likely age is ${\sim}500\textrm{--}800$~Myr.
    \label{fig:overdensity_colour}
	}
\end{figure*}

We further quantify this distance constraint by investigating the 
optical--NIR
colour of the overdensity.
The \textit{HST} F814W GALFIT residual image was PSF matched by convolving it with the \textit{Spitzer} IRAC1 PSF.
This low resolution PSF (full width at half maximum $\sim 2\arcsec$) 
was interpolated to the \textit{HST}/ACS pixel scale using the \texttt{photutils} package.
We then used the \texttt{reproject} package\footnote{\url{https://reproject.readthedocs.io}} to align and reproject the PSF-matched F814W residual image onto the IRAC1 pixel grid.
We repeated this procedure for the F606W image.
The same $7^{\prime\prime}$ diameter circular aperture was then used to measure the $\mathrm{F814W} - \mathrm{IRAC1}$ colour on the F814W and IRAC1 residual images.
We find $\mathrm{F814W} - \mathrm{IRAC1} = -1.5 \pm 0.2$.
The uncertainty includes error in the sky value, pixel errors and GALFIT model uncertainties in both bands, the last of which are likely underestimated.
In the left panel of Figure \ref{fig:overdensity_colour}, the colour of the overdensity is marked as the light blue square.

For reference, we also plot the colours of extended and point-like sources in Figure~\ref{fig:overdensity_colour}.
These colours were measured on the PSF-matched \textit{HST}/ACS and \textit{Spitzer}/IRAC images in 6 pixel diameter circular apertures.
Sources were then matched back to the \textit{HST} catalog to use the \textit{HST} concentration $C_{3-7}$ to separate extended from point-like sources.
The overdensity appears too faint in the NIR to be consistent with a background galaxy in projection.

If the overdensity is at roughly the same distance as DGSAT~I (embedded in the
galaxy or nearby but superimposed in projection), then the luminosity is $L_V \sim 10^7~L_\odot$ 
and the physical diameter is $\sim 1$~kpc.
These values are comparable to a low-luminosity dwarf irregular (dIrr) galaxy --  
in this case one with a peculiar, cross-like morphology.
The narrow-band H$\alpha$ photometry from \citet{MD2016} resulted in a non-detection that implies quiescence for DGSAT~I overall, but would not exclude a lower-mass galaxy with a normal
star formation rate.
Similarly, the non-detection of cold gas by \citet{papastergis2017},
with a 5$\sigma$ upper limit of $M_{\rm HI} < 2.4\times10^8 M_\odot$, is an important constraint
for DGSAT~I itself but not for a lower-mass dwarf.
We turn instead to an analysis of how likely it is for
a dIrr galaxy in the Pisces--Perseus supercluster to be randomly superimposed on DGSAT~I.
The luminosity function of galaxies in the field from the ALFALFA survey is
$\phi \sim 0.05~\mathrm{Mpc}^{-3}~\mathrm{dex}^{-1}$ for $M_{\rm HI} \sim 10^8~ M_\odot$ \citep{jones2018}.
We assume a very generous depth of the filament of 20 Mpc and that the density in the filament can be no more than ${\sim}10$ times greater than in the field.
Then, adopting a $10~\mathrm{kpc} \times 10~\mathrm{kpc}$ footprint of DGSAT~I, 
the chance of such a projection is 
$10 \times \phi \times (0.01 \times 0.01 \times 20)~\mathrm{Mpc}^3 \sim 0.001$ (0.1\%)
at best, given that 
our filament density and depth assumptions are almost certainly overestimates.

\cite{MN2019} measured a total metal abundance from spectroscopy of DGSAT~I to be
$[\mathrm{M}/\mathrm{H}] = -1.8 \pm 0.4~\mathrm{dex}$ and a mass-weighted age of
$8.1 \pm 0.4~\mathrm{Gyr}$.
We take these values to be lower and upper limits on metallicity and age,
respectively, for the overdensity, which we assume formed more recently than
the rest of the galaxy.
In the right panel of Figure \ref{fig:overdensity_colour}, we plot 
age against
$\mathrm{F606W} - \mathrm{F814W}$ colour for SSPs
spanning a broad range of metallicities created using FSPS \citep{conroy2009,
conroy2010}.
The observation for DGSAT~I is plotted as the black error bars,
and is consistent with an old, metal-poor population
as found by spectroscopy.
The central overdensity colour is marked with a dash-dotted vertical line:
its blue color cannot be reproduced by a metallicity effect, and requires an
age of ${\sim}1.5~\mathrm{Gyr}$ or younger.
If we adopt a reasonable assumption that the metallicity is greater than
$\log(Z/Z_\odot) \sim -1.5$,
then an age range of ${\sim}500\textrm{--}800$ Myr is preferred.
The SSPs also provide a mass estimate for the overdensity of $(5.5 \pm 0.4)
\times 10^6~M_\odot$, 
which is ${\sim}4\%$ of the stellar mass of the galaxy.

This small component of recent localised star formation is intriguingly similar to the spectroscopic
finding of an extended star formation history (SFH) for the bulk of the galaxy,
with ${\sim}25\%$ of the stars formed within the last 3 Gyr, and
${\sim}5\%$ within the last 400 Myr (\citealt{MN2019};
note their Figure~A3 shows that the instrument footprint did not include the blue overdensity).
On the other hand, it has also been shown that 
biases in spectroscopic modelling can lead to artificially extended SFHs
\citep{webb2022},
and hence it is possible that the galaxy is uniformly old, except for
this one obvious region of recent star formation.
Reconciling the younger blue overdensity within a much older UDG may have
implications for understanding the formation of DGSAT~I and UDGs in general,
which will be discussed in Section~\ref{sec:backsplash}.

\subsection{The Globular Cluster System of DGSAT~I}
\label{sec:GCs}

We now consider several different properties of the GC system of DGSAT~I, including their
spatial distribution, total numbers and masses, luminosity function and colours.
In Figure \ref{fig:cmd}, the right panel shows the GC candidates in the
vicinity of DGSAT~I marked with small, light blue circles. 
The dashed outer ellipse is the half-light boundary of DGSAT~I ($R_{\rm e}$;
Section~\ref{sec:galfitresults}) and the solid inner ellipse is the half-number boundary of the
GC system (derived below).
There is clearly an overdensity of GCs within $1 R_{\rm e}$, and relatively few farther out, especially considering that these will include contaminant objects.

\begin{figure}
	\includegraphics[width=0.50\textwidth]{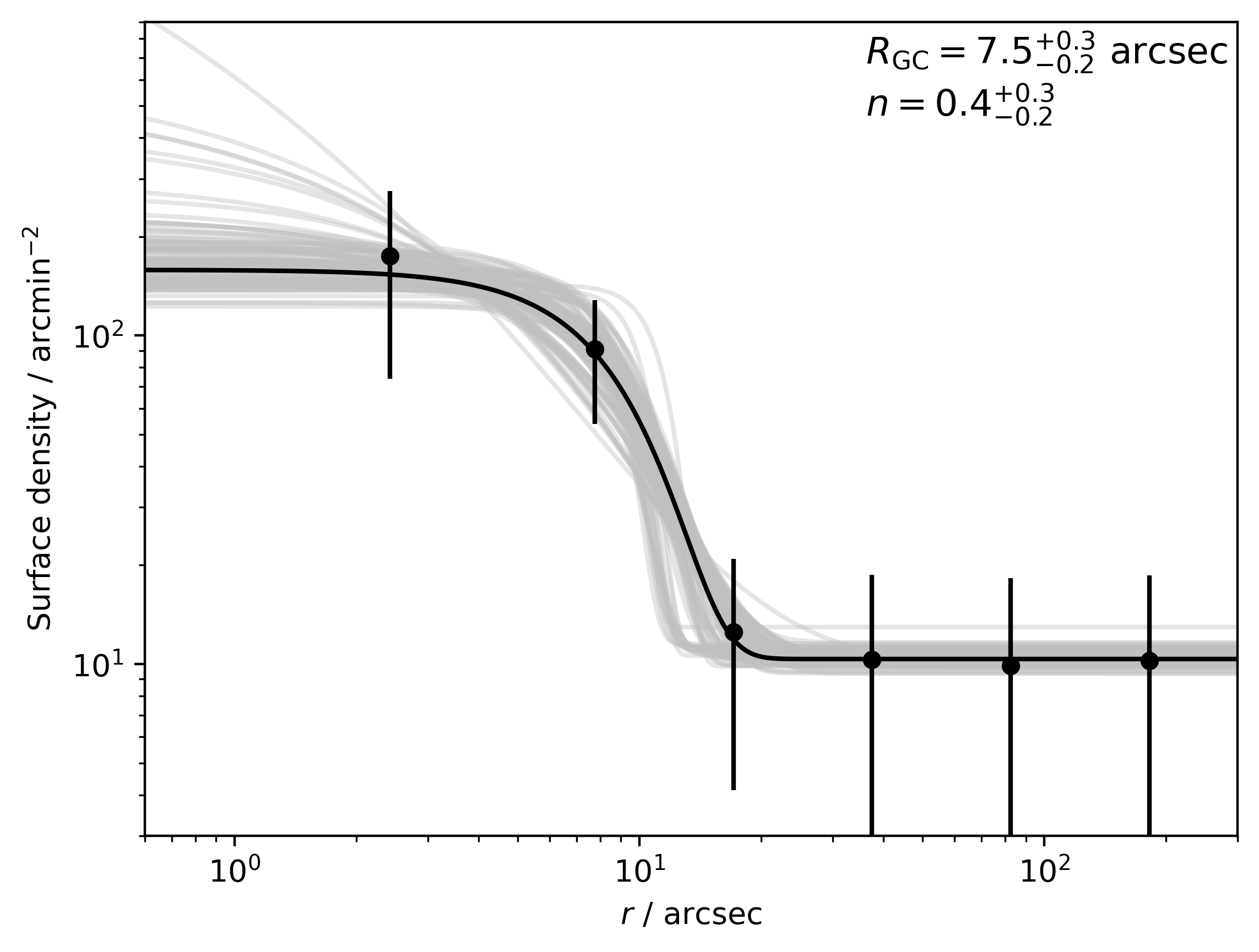}
	\caption{
    GC radial density profile along the semi-major axis centred on DGSAT~I. 
    Shown in light grey are one
    dimensional S\'{e}rsic model fits to 100 different samplings of the radial
    distribution by randomly drawing the bin locations and sizes from uniform
    distributions.
    At small radii, the error bars are Poisson errors from the number of
    counted GC candidates.
    At large radii ($r > 12\arcsec$), the error bars are instead the measured
    scatter of the number of point sources that satisfy our GC cuts in 100
    DGSAT~I-sized elliptical apertures placed randomly around the image.
    \label{fig:gc_density}
	}
\end{figure}

The half-number radius of the GC system was measured by fitting S\'{e}rsic profiles
to the radial surface density of GCs around DGSAT~I. 
The surface density was measured in elliptical annuli with position angles and
ellipticities fixed to those of the UDG.
A suite of 100 surface density profiles were created by randomly varying both the
widths and radial locations
of the elliptical annuli, and
then fitting a one dimensional S\'{e}rsic profile plus a constant background.
The surface density was measured out to at least $25 R_\mathrm{e}$ to ensure the
background level was reached in all profiles.
In Figure \ref{fig:gc_density}, the 1D S\'{e}rsic plus background fits to the 100
measurements of the surface density profile are plotted in grey, the black
points show one such measurement, and the black solid curve is the adopted
model based on the median values of the model parameters.
At small radii, the error bars reflect Poisson statistics.
At large radii ($r > 12\arcsec$), the error is instead 
driven by fluctuations in background counts, which we estimate using
the standard deviation
in the number of GC candidates counted in 100 
DGSAT~I-sized elliptical apertures placed randomly around the field.

We find a total number of detectable GCs around DGSAT~I of $12.0^{+0.3}_{-0.6}$, a
GC system half-number radius of $R_\mathrm{GC} = 7.5^{+0.3}_{-0.2}~\mathrm{arcsec} =
2.68^{+0.11}_{-0.07}~\mathrm{kpc}$ and a S\'{e}rsic index $n =
0.4^{+0.3}_{-0.2}$.
The size of the GC system compared to the size of the galaxy is
$R_\mathrm{GC}/R_\mathrm{e} = 0.67 \pm 0.03$.
To estimate the total number of GCs, we must correct for incompleteness effects as a function of magnitude, which we return to after discussion of the GC luminosity function (GCLF).

\begin{figure*}
    \includegraphics[width=0.99\textwidth]{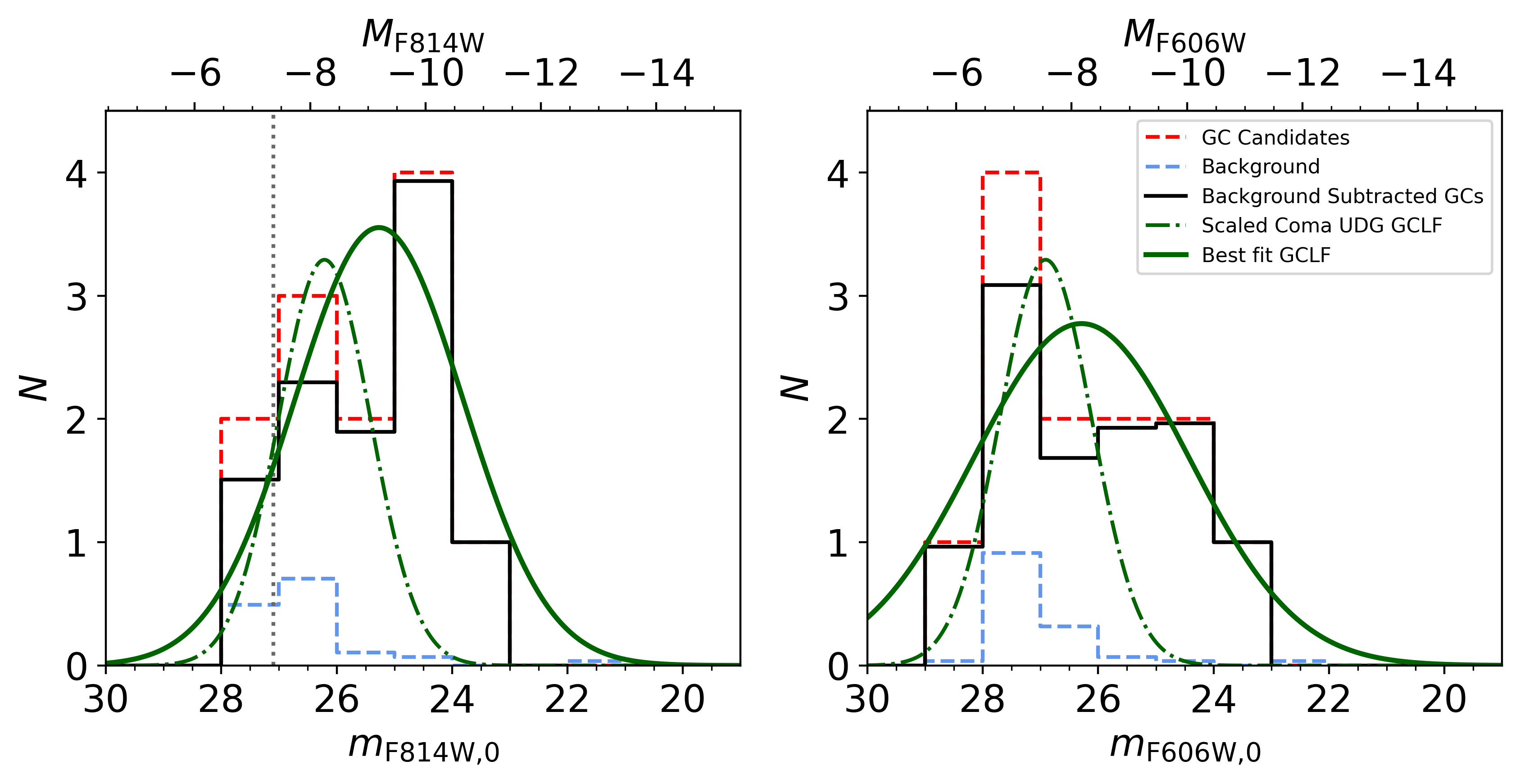}
	\caption{
    The GCLF of DGSAT~I in F814W (left) and F606W (right).
    The raw counts of GC candidates within an aperture of size $1.5 \times R_\mathrm{e}$ are shown as the red dashed histograms.
    The dashed blue histogram shows the expected contribution from
    background sources.
    The background corrected GCLF is shown in black.
    The 50\% completeness limit in F814W is shown with a vertical dotted line.
    The dash-dotted green curve shows
    a Gaussian GCLF with a shape similar to that observed in Coma UDGs \protect\citep{saifollahi2022}
    with a peak at $M^0_\mathrm{F814W} = -8.25$ ($M^0_\mathrm{F606W} = -7.55$), a width $\sigma = 0.8$ and a total of $7$ GCs (scaled to match the observed number of GCs near the peak and
    corresponding to a specific frequency $S_N \simeq 3$).
    The typical GCLF for cluster dwarfs would be even narrower.
    The Gaussian GCLF fit to the observations (solid green line) is both brighter ($M^0_\mathrm{F814W} = -9.2$, $M^0_\mathrm{F606W} = -8.2$) and significantly broader ($\sigma_\mathrm{F814W} = 1.5$, $\sigma_\mathrm{F606W} = 1.9$), containing a total of ${\sim}12$ GCs for a specific frequency $S_N \simeq 6$. 
    \label{fig:gclf}
	}
\end{figure*}

Figure \ref{fig:gclf} shows the GCLF of DGSAT~I in both F814W and F606W,
with apparent and absolute magnitudes on the lower and upper axes, respectively.
The dashed red histograms are the raw GC candidate counts inside an aperture of size
$1.5 \times R_\mathrm{e}$.
Blue dashed histograms are the expected background contributions. 
These were measured for point sources that satisfy our colour and magnitude cuts, 
in an annulus with inner radius $6 \times R_\mathrm{e}$ and outer radius $10 \times R_\mathrm{e}$, and
scaled to match the area of the GC candidate aperture.
In solid black is the GCLF after correcting for background contamination.
No correction is made here for completeness effects, but as a reminder, the 50\% completeness limit is at F814W = 27.1
(Section~\ref{sec:phot}).
The F606W cutoff is at 27.9 mag, which is due not to direct incompleteness but to
the need to have matched detections in the F814W band within the adopted range of GC colours.

For context, the GCLF peaks or turnover magnitudes, $M^0$,
from the literature on galaxies from giants to dwarfs are as follows.
The metal-poor and metal-rich GC subpopulations in M87 have
$M^0_\mathrm{F814W} = -8.4$ and $-7.9$, respectively \citep{peng2009}.
For dEs in the Virgo cluster, $M^0 = -8.25$ on average in F814W,
and $-7.55$ in F606W (\citealt{miller2007}; after updating the distance to Virgo).
A comprehensive study of Virgo and Fornax galaxies found $M^0$ ranging from $\simeq -8.6$
for giants to $\simeq -8.3$ for dwarfs in the F850LP band (which is very close to F814W), with
an intrinsic galaxy-to-galaxy scatter 
of $\simeq 0.2$~mag \citep{villegas2010}.
Therefore the plausible ranges for $M^0$ based on normal cluster dwarfs
are between $\sim -7.9$ and $-8.6$ in F814W, and $\sim -7.2$ and $-7.9$ in F606W.

The observed GCLFs for DGSAT~I in Figure~\ref{fig:gclf} appear to be brighter than these expectations,
particularly in the F814W band.
We carry out fits to our data assuming Gaussian distributions, finding
mean and dispersion values of 
$\mu = 25.3 \pm 0.7$, $\sigma = 1.5 \pm 0.5$ in F814W (corresponding to $M^0 = -9.2 \pm 0.8$),
and $\mu = 26.3 \pm 0.8$, $\sigma = 1.9 \pm 0.7$ ($M^0 = -8.2 \pm 0.9$) in F606W.
Although formally these $M^0$ values are consistent with expectations, given the scatter and
uncertainties, it should be kept in mind that the fitted magnitude spreads are abnormally large,
compared to a typical $\sigma \sim 0.6$~mag for cluster dwarfs \citep{villegas2010}.

Alternatively, we can focus on constraints from the very bright GCs in DGSAT~I
(five of them with $\mathrm{F814W} < 24.8$).  Given a GCLF with parameters ($M^0,\sigma$),
we simulate mock datasets of ten GCs at a time, brighter than the 80\%
completeness limit of F814W~$=26.9$.
Then we check what fraction of these datasets yield 5 bright GCs.
We find that the fraction is effectively zero for typical dwarf GCLFs,
and rises no higher than $\sim$~10\% for even the brightest and broadest GCLFs
known in Virgo/Fornax cluster giant galaxies.
Thus we have a definitive detection of overluminous GCs in DGSAT~I.

We will discuss the implications of the anomalous GCLF in the next Section, but
here briefly consider whether or not the adopted distance of 78~Mpc could be a factor in this result.
DGSAT~I may be associated with a group at $\sim$~71~Mpc (discussed later), but
this would imply a GCLF only $\sim$~0.2 mag fainter.
Accommodating a canonical value of $M^0$ would require a $\sim 1.0$~mag shift,
i.e.\ a distance of
$\sim 50$~Mpc.
The redshift-based distance should not be that far off:
the allowed range based on calculations with different Hubble Flow models in NED
is 73--90~Mpc.
Also, a closer galaxy would be intrinsically smaller and fainter, which would 
make the measured high velocity dispersion and red GC colours more problematic (discussed later).
Furthermore, the large spread in GC luminosities is also anomalous, and
would not be resolved by a distance change.

We return to a total estimate of GC numbers.
Given the relative brightness of the GCLF, corrections for incompleteness are small,
leading to a total number of $N_{\rm GC} = 12 \pm 2$,
with the uncertainty from contamination and incompleteness corrections.
The implied GC specific frequency, defined as 
$S_N = N_{\rm GC} 10^{0.4(M_V + 15)}$, is
$5.2 \pm 0.9$. 

An alternative way to characterize the GC content is by the total mass,
$M_{\rm GC}$, which is less sensitive than $N_{\mathrm GC}$ to
incompleteness corrections and to the assumed GCLF.
For our GC detections, their total luminosity in F814W is
$L = 5.0\times10^6 L_\odot$ (after completeness and contamination corrections),
and adopting $M/L \simeq 1.33$ again (Section~\ref{sec:galfitresults}),
we find $M_{\rm GC} \simeq 6.6\times10^6 M_\odot$.
The mass fraction of GCs relative to the galaxy is then 
$M_{\rm GC}/M_\star \simeq 0.020$.

As shown in Figure~\ref{fig:cmd}, the GCs have a remarkably narrow range of colour that is also very similar to the colour of the galaxy light (marked by a vertical gold line).
Quantitatively, the uncertainty-weighted mean for the colour of the GCs brighter than the turnover magnitude is
$\mathrm{F606W} - \mathrm{F814W} = 0.46$ mag, 
with a standard deviation of 0.05 mag.
The estimated typical errors of the colours are $\sim$0.05 mag, so the GCs are consistent with having zero colour spread.

The small scatter in the GC colours will be interpreted in the next Section, 
but also provides a novel constraint on internal dust in UDGs.
If the GCs are embedded randomly within the galaxy along the line of sight, then they should show differential reddening from any diffusely distributed dust.
The colour spread above implies an approximate upper limit of $A_V \sim 0.2$, ruling out some
of the preferred dust solutions from fitting of the spectral energy distribution of DGSAT~I \citep{pandya2018,buzzo2022}.
and reinforcing suspicions of certain systematic problems with stellar population synthesis models.

\section{Discussion}\label{sec:disc}

Here we discuss the results presented above, and attempt to 
interpret them with
a coherent picture of the formation history of DGSAT~I, its GCs and implications for UDGs in general.

\subsection{Is DGSAT~I a backsplash galaxy?}
\label{sec:backsplash}

As discussed in \citet{MD2016}, DGSAT~I is located within a filament of the 
Pisces--Perseus supercluster,
and in apparent proximity to a massive galaxy group (or poor cluster) 
at 1.3~deg with a velocity separation of ${\sim}500~\mathrm{km}~\mathrm{s}^{-1}$.
This group is centred near the galaxy NGC~507\footnote{There is a ``Zwicky cluster'' called
Zw 0107+3212 or UGCl020
that appears to actually be an unbound association of galaxies and groups along the filament,
including the NGC~383 and NGC~507 groups
(see positional maps in \citealt{sakai1994}).
The NGC~507 group is also called WBL 503, PCC S34-113 and MCXC J0123.6+3315.} 
at a line-of-sight distance of ${\sim}71$ Mpc, 
which would imply a projected distance to DGSAT~I of ${\sim}1.6$~Mpc.
Given a velocity dispersion and virial radius for this group of
$\sigma \sim 500$~km~s$^{-1}$ and $R_{\rm vir} \sim 1.1$~Mpc, respectively
\citep{crook2008},
a projected-phase space diagram can be used to diagnose the potential association of DGSAT~I with the group.
Using the simulated diagrams of \citet{rhee2017}, DGSAT~I is very likely to be
either an unassociated ``interloper,''
or on first infall to the cluster with little tidal mass loss
(see also Figure~5 from \citealt{joshi2021}).
There is a ${\sim}10\%$ chance that it is a backsplash galaxy around apocenter after one or more pericentric passages.

Although this backsplash probability is low, DGSAT~I may also be a relatively rare object,
so backsplash is in principle a tidy explanation.
The young, blue overdensity initially appears to be a problem here, since
the simulations of \citet{benavides2021} found
backsplash UDGs to be stripped of all gas from ram pressure during their interaction
(see also \citealt{borrow2022}),
leading to rapid quenching\footnote{
A caveat about the 
\citet{benavides2021} simulations is that they did
not actually  produce any red field galaxies with a similar
mass and size to DGSAT~I, predicting $R_{\rm e} \sim 2.5$~kpc or smaller in its mass range.
There is a similar inability of the simulations to reproduce the large sizes
of UDGs in clusters,
suggesting that there may be an important channel for UDG formation that is entirely missed.}.
The time to travel ${\sim}1\textrm{--}1.5$ Mpc after pericenter is ${\sim}2$ Gyr, so
there appears to be a mismatch between the timescales for splashback and quenching (${\sim}0.5$ Gyr)
for DGSAT~I.
On the other hand, quenching processes and timescales 
for low-mass galaxies in massive groups and clusters are not well understood,
and DGSAT~I could have taken up to a few Gyr to quench after pericentric passage
\citep{wetzel2013,oman2021}.

Another possibility is
that DGSAT~I is on first infall to the NGC~507 group,
with its cold gas supply being cut off as it approaches,
and the final episode of star formation occurring ${\sim}500$~Myr ago.
Galaxies are not thought to quench this early in the infall process,
but perhaps DGSAT~I had a very feeble gas reservoir to begin with, which
was quickly exhausted after entering the outskirts of the group.

A final possibility is that DGSAT~I is truly a ``field'' galaxy that recently
quenched in a low-density
environment without a conventional environmental trigger.
Some simulations produce UDGs with bursty cycles of star formation and
quiescent episodes in between \citep{dicintio2017}, but presumably there should still
be cold gas lurking in the vicinity, which was not detected in the
case of DGSAT~I \citep{papastergis2017}.
A combination of this feedback scenario with first infall (above) to remove
the gas might work.
Alternatively, there are proposed mechanisms for ram pressure stripping
and shock heating in the intergalactic medium that could 
quench dwarf galaxies in the field
\citep{benitez-llambay2013,pasha2022,yang2022}.
A more remarkable possibility would be that DGSAT~I has a massive enough dark matter halo (see next Sections) to behave like a more luminous early-type galaxy in its ability to continue with low-level star formation--whether via direct gas accretion, minor mergers or gas condensation out of a warm interstellar medium
(e.g.\ \citealt{young2014}).

\subsection{Globular cluster spatial distribution, numbers and connection to halo mass}
\label{sec:GCdisc}

We now begin to examine how the properties of the GC system of DGSAT~I 
relate to other galaxies and to potential formation scenarios.
We found in Section~\ref{sec:GCs} that $R_\mathrm{GC}/R_\mathrm{e} \sim 0.7$.
DGSAT~I therefore joins with an emerging picture where
UDGs typically have GC systems more compact than, or similar to, their stellar light
\citep{saifollahi2022,danieli2022}--in contrast to other dwarf galaxies where the
GC systems are more extended (e.g.\ \citealt{lim2018,carlsten2022}).
This difference may be interpreted by first clarifying why GC systems are relatively extended in normal dwarfs.
Unlike massive galaxies, 
dwarfs are expected to have assembled their stars and GCs almost entirely in-situ rather than from accretion events
(e.g.\ \citealt{choksi2019}).
An extended old, metal-poor component of stars and GCs can then be traced to an ``outside-in'' formation
process, where gas collects at later times in the central regions and forms younger, enriched stars there,
with feedback pushing the older stars to larger radii
\citep{graus2019,mercado2021,emami2021,kado-fong2022}.
The central number density of the GCs may then be further reduced by their tidal destruction in these dense regions.
GCs may also form preferentially toward the outskirts of dwarf galaxies
in early, chaotic conditions \citep{sameie2022}.

For UDGs to have GC distributions that are similar to the field stars, or even more centrally concentrated,
suggests that they have evolved differently from normal dwarfs.
For example, if UDGs are normal dwarfs that were tidally heated, 
then the large $R_{\rm GC}/R_{\rm e}$ ratio
should be preserved by the expansion process \citep{saifollahi2022}.
Similar arguments can be made for mergers or feedback-driven expansion, 
unless perhaps there was only one burst of star formation where the GCs
formed preferentially in the central regions.
Such a scenario was presented by \citet{danieli2022} in the context of NGC~5846-UDG1,
where high gas densities at early times
triggered unusually efficient GC formation along with strong gas outflows that expanded the galaxy and truncated star formation.
In a similar vein,
\citet{villaume2022} argued for an ``inside-out'' formation history for the iconic Coma UDG Dragonfly 44 (DF44),
based on unusual radial gradients in its field stars, with more GC-like abundance patterns in its center
(relatively iron-poor and magnesium-enhanced, and presumably older).
There is no evidence for this abundance pattern in DGSAT~I \citep{MN2019}, 
although the spectroscopic data are not deep enough to detect such a gradient clearly,
and in any case the clump of recent central star formation (Section~\ref{sec:overdensity})
demonstrates that this galaxy did not form completely inside-out.

We next place the numbers $N_{\rm GC} \sim 12$ and total mass $M_{\rm GCS} \sim 6\times10^6 M_\odot$
in the GC system of DGSAT~I (Section~\ref{sec:GCs}) in context,
keeping in mind that these parameters are not trivially interchangeable,
owing to the unusual GCLF.
The empirical scaling relation between $N_{\rm GC}$ and
halo virial mass $M_{\rm vir}$ \citep[e.g.][]{burkert2020}
suggests that this UDG has $M_{\rm vir} \sim 7 \times 10^{10} M_\odot$.
On the other hand, if the GC system mass is more fundamental,
then the implied halo mass is $M_{\rm vir} \sim 2 \times 10^{11} M_\odot$
\citep{harris2017}.
For a DGSAT~I stellar mass of $M_\star \simeq 3 \times 10^8 M_\odot$
(Section~\ref{sec:galfitresults}),
standard expectations for halo mass are in the range ${\sim}(0.6 \textrm{--} 1.4)~\times 10^{11}~M_\odot$
\citep{rodriguez-puebla2017},
so $N_{\rm GC}$ implies a normal halo, 
while $M_{\rm GC}$ implies an overmassive one.
The latter result supports the classification of DGSAT~I as
a ``failed galaxy''--although it is not an extreme example,
which we know in any case because of its extended SFH.

\begin{figure}
	\includegraphics[width=0.50\textwidth]{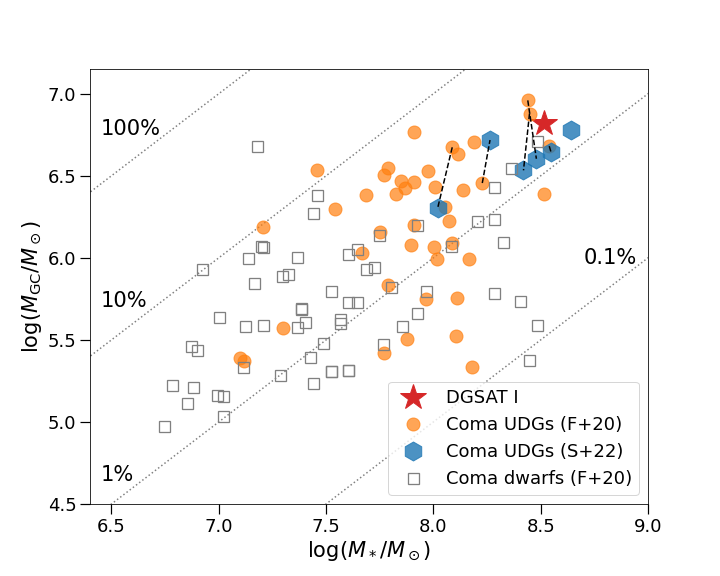}
	\caption{
    GC system mass versus host-galaxy stellar mass.
    DGSAT~I is shown by a large red star, 
    Coma UDGs from \protect\citet{saifollahi2022} by blue hexagons,
    Coma UDGs from \protect\citet{forbes2020} by small orange circles,
    and other Coma dwarfs from \protect\cite{forbes2020} by small grey open squares.
    Note that five of the high-mass Coma UDGs are duplicated between the two data-sets,
    and are shown twice connected by dashed black lines to illustrate the systematic differences between the studies.
    Diagonal grey dotted lines show GC-to-stellar mass fractions $f_{\rm GC}$ as labelled.
    In terms of GC mass, DGSAT~I appears to be a ``field'' counterpart to Coma cluster UDGs.
    \label{fig:gc_mass}
	}
\end{figure}

The most extensively studied samples of GC systems around UDGs
are in the Coma cluster, with the largest compilation presented
in \citet{forbes2020}, along with GC data on other Coma dwarfs.
We show the masses of these galaxies' GC systems versus their total stellar mass
in Figure~\ref{fig:gc_mass},
where we have adopted the same $M/L$ as for DGSAT~I,
and mean GC 
masses of $(0.8$--$1.2) \times 10^5 M_\odot$
following the \citet{harris2013} scaling relation with host galaxy mass.
At masses of $M_\star \sim 10^8 M_\odot$, the Coma UDGs (orange circles) have 
GC mass fractions of $f_{\rm GC} \equiv M_{\rm GC}/M_\star \sim$~1--4\%,
while the other Coma dwarfs (grey squares) have lower values of $f_{\rm GC} \sim$~0.5--1.5\%
(see \citealt{lim2018}).
DGSAT~I appears intermediate to these populations, with $f_{\rm GC} \simeq$~2\%,
although we caution that the analysis methods were not homogeneous.

More detailed analysis of Coma UDGs, using methods
fairly similar to ours used here for DGSAT~I, has revised their
GC numbers downward substantially
(\citealt{saifollahi2022}; it is beyond the scope of this paper to establish
the reasons for the differences in GC counts between various studies).
We plot their results as blue hexagons in Figure~\ref{fig:gc_mass},
where we have adopted a mean GC mass of $2\times10^5 M_\odot$ from integrating their combined GCLF (which is overluminous compared to normal cluster dwarfs),
and find $f_{\rm GC} \sim$~1--3\%. 
DGSAT~I thus appears similar in GC system mass to these Coma UDGs, 
and is consistent with being an analogue found in a lower-density environment.
Furthermore, if DGSAT~I had fallen in to a cluster $\sim$~8 Gyr ago and quenched before
forming half of its present-day stellar mass, it would appear 
close to the Coma UDG DF17 in Figure~\ref{fig:gc_mass} 
(blue hexagon to the left of the red star).

Another important point of comparison is to the GC systems of UDGs in lower density
environments.  The data so far suggest lower GC numbers on average in galaxy groups
and in lower-mass galaxy clusters
\citep{prole2019,lim2020,somalwar2020,marleau2021}--similar to the environmental trends for normal dwarfs (e.g.\ \citealt{carlsten2022}).
 These trends are generally attributed to higher cluster formation efficiency
 in higher density environments (e.g.\ \citealt{peng2008}), and it is expected
 that field UDGs, and field dwarfs in general, would have the most sparsely populated GC systems.
DGSAT~I is thus an outlier, as a GC-rich galaxy in a low-density environment--implying
either that it was born in a high-density environment and then ejected
(while somehow managing to continue star formation for many Gyr),
or that it had a different formation pathway than typical for dwarfs.

An additional related property of DGSAT~I is its stellar velocity dispersion,
which was measured to be $\sigma = 56 \pm 10~\mathrm{km}~\mathrm{s}^{-1}$ by \citet{MN2019}:
the highest value known for any UDG.
This measurement can be converted to a dynamical mass estimate within the
three-dimensional half-light radius
of $M_{1/2}(< 4.9~\mathrm{kpc}) = (1.1 \pm 0.4) \times10^{10} M_\odot$ \citep{wolf2010}, implying a dark matter fraction of 95--98\%
and a mass-to-light ratio of $M/L_V = 56 \pm 20$.
This high dynamical mass is ${\sim}10\textrm{--}20$ times higher than that of star-forming field UDGs
of similar stellar mass \citep{kong2022} and implies
a high halo mass of
$M_{\rm vir} \sim 3\times10^{11}~M_\odot$ or higher \citep[e.g.][]{forbes2021}.
It also leads to the expectation of a very populous GC system--in agreement with our observations here.
More specifically, there is an empirical correlation between $M_{1/2}$ and $N_{\rm GC}$ (or $M_{\rm GCS})$
for ``normal'' galaxies, which so far appears to be followed by UDGs as well
\citep{harris2013,vandokkum2017,toloba2018,gannon2020}. 
DGSAT~I fits in with this trend, based on its $M_{\rm GCS}$ rather than its $N_{\rm GC}$.

\subsection{Globular cluster colours, luminosities and formation implications}
\label{sec:GCcol}

Most of the attention to GCs in UDGs has revolved around their numbers
and luminosity functions.  A few comparisons of GC and host
galaxy {\it colours} have been made, but there has been no systematic attempt
to understand these.
This aspect is of fundamental importance since there are well-known colour and metallicity
trends in normal dwarfs that are closely connected to their underlying
assembly histories.
More luminous galaxies are more metal-rich on average,
likely as a consequence of their deeper potential wells 
retaining metals more easily.
For quiescent galaxies, this leads to redder colours with brighter magnitudes. 
GC colours and metallicities also follow the same correlation with
host galaxy luminosity, but with an offset.
In all types of galaxies studied to date, the mean GC colours are
generically lower, and their metallicities lower,
than those of their host galaxies \citep[e.g.][]{lotz2004,peng2006,larsen2014}, 
probably as a reflection of early star
formation bursts that were conducive to GC formation, followed by lower rates
of continuous star formation that built up the bulk of the galaxy's field
stars.
This generic offset is smaller for dwarfs than for giants, but still present, and provides a
critical and novel avenue for comparing the formation histories of UDGs and
normal dwarfs.

\begin{figure}
	\includegraphics[width=0.50\textwidth]{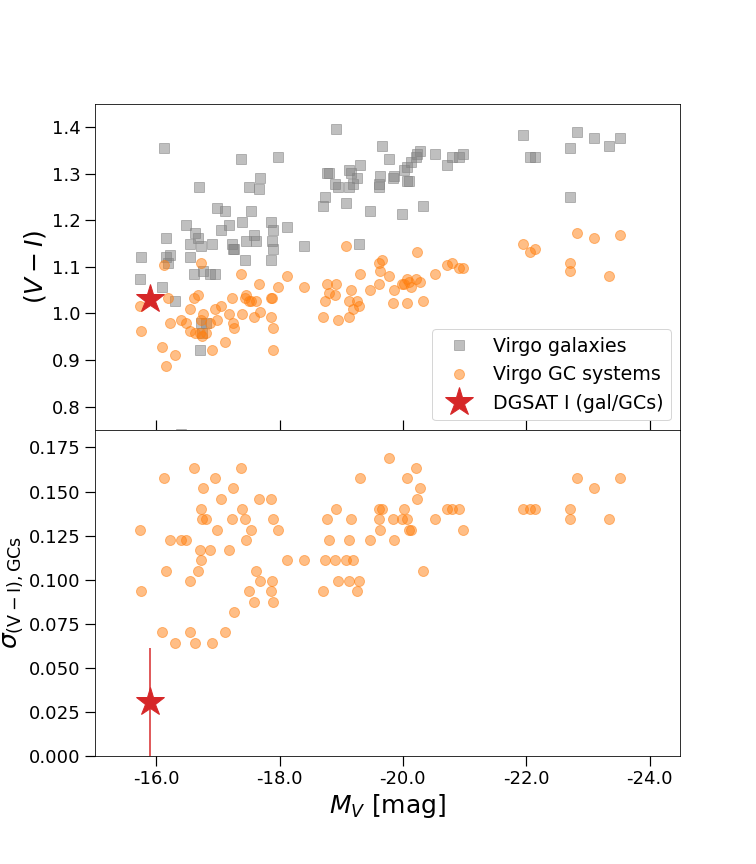}
	\caption{
    Colour trends for galaxies and their GC systems as a function of host galaxy absolute magnitude.
    {\it Top}:  Mean colours, with Virgo galaxies and GCs systems shown as grey squares and orange circles, respectively.
    The large red star shows DGSAT~I, with identical colours for the galaxy and its GCs.
    {\it Bottom}: Colour spreads of the GC systems, with symbols as in the top panel.
    DGSAT~I diverges from the cluster dwarf trends, with a colour spread consistent with zero.
    This suggests that the GCs have nearly identical ages and metallicities, and formed in the same starburst event.
    \label{fig:gc_colors}
	}
\end{figure}

In order to compare results across various studies, we need to put the colours on a common scale.
For the DGSAT~I galaxy and GCs, the colour measurements of
F606W$-$F814W = 0.46 (AB) in both cases 
(Sections~\ref{sec:galfitresults} and \ref{sec:GCs}) 
correspond to $V-I = 1.03$ (Vega, using the \citealt{saha2011} transformations).
A comparison sample of Virgo cluster galaxies is taken from the ACSVCS survey \citep{peng2006,ferrarese2006},
with a conversion from $g_{475}-z_{850}$ to $V-I$ based on transformations from \citet{usher2012}.
The results are shown in the top panel of Figure~\ref{fig:gc_colors},
with colours for GC systems and their host galaxies plotted separately,
as a function of galaxy absolute magnitude.  
DGSAT~I is bluer than average for the cluster dwarfs, probably as a result of its low metallicity rather than an age effect (cf.\ left panel of Figure~\ref{fig:overdensity_colour}).

More remarkable is the {\it red} colour of the DGSAT~I GCs relative to cluster dwarf GCs, which is typical of a galaxy ${\sim}10$ times more luminous, and could be indicative of an overmassive dark matter halo, taking redness as a proxy for metallicity and potential well depth
(see also similar findings for GCs around luminous Virgo UDGs in \citealt{lim2020}).
Some other property of the GCs could also be responsible,
as it has been found that GC colour--metallicity relations are {\it not} universal from galaxy to galaxy,
particularly for metal-poor GCs \citep{usher2015,powalka2016,villaume2019,fahrion2020}.
Such an interpretation is plausible, given the very low spectroscopic metallicity of the stars in DGSAT~I
([M/H] $\sim -1.8$), which contrasts with the [$Z$/H]~$\simeq -0.95$ for the GCs implied by
a standard colour--metallicity relation \citep{usher2012}.
The most likely underlying explanation in this case would be that the GCs in DGSAT~I are more $\alpha$-enhanced (like their host galaxy) and older than the cluster dwarf GCs
(see also related discussions in \citealt{sharina2005,fahrion2020}).

The close agreement of the galaxy and GC system colour in DGSAT~I is almost unprecedented among the cluster dwarfs.  
In principle, this could imply coevolution of the GCs and field stars, although the galaxy's extended SFH suggests that there is instead a coincidental combination of different ages and metallicities that have led to the same colour (vertical downward evolution in the left panel of Figure~\ref{fig:overdensity_colour}).

The bottom panel of Figure~\ref{fig:gc_colors} shows the colour spreads of the same GC systems as in the top panel.
Dwarf galaxy GC systems have narrower colour spreads on average than in giants, primarily because they are purely metal-poor systems missing the metal-rich GC subpopulations.
DGSAT~I 
stands out here with its particularly narrow colour spread that 
is consistent with zero.
The implication is that the GCs may have nearly identical ages and metallicities,
and formed in a single starburst event.
A similarly narrow colour distribution of metal-poor GCs was found in NGC~5846-UDG1
\citep{muller2020,muller2021,danieli2022}, consistent with the interpretation of this galaxy,
with its extremely high $S_N$, 
as forming in a short-lived burst of early, clumpy star formation, akin to a failed galaxy.
The same monochromatic GCs phenomenon was found in 
the dark-matter-poor dwarfs NGC~1052-DF2 and -DF4 
\citep{vandokkum2022b}, where the single starburst is thought to reflect formation in a high-speed galactic collision \citep{vandokkum2022a}.

A more systematic understanding of colour trends in UDGs is important, 
and in principle a good {\it HST} data set is available for six galaxies in 
the Coma cluster
\citep{beasley2016b,vandokkum2017,saifollahi2022}.
Unfortunately, comparisons of galaxy and GC colour results for the same
systems across these different studies reveal inconsistencies that
are beyond the scope of this paper to sort out.
Qualitatively, these UDGs appear similar to DGSAT~I in
hosting GCs that are redder than expected.
On the other hand, most of these Coma UDGs appear to have substantial GC colour
spreads in agreement with the trend for ``normal'' cluster dwarfs in
Figure~\ref{fig:gc_colors}.
One exception is DF07, where the brighter GCs exhibit a very narrow colour spread,
similar to DGSAT~I (see \citealt{saifollahi2022} Figure~A7).

Turning now to GC luminosities,
a near universal GCLF that is observed across large samples of galaxies is thought to be a reflection of fundamental processes of star cluster formation in a power-law mass function, followed by evolution through internal relaxation and tidal effects
(e.g.\ \citealt{fall2001}),
although it remains a challenge to model GCLFs in detail
(e.g.\ \citealt{li2019,reina-campos2022,rodriguez2022}).
There are mild correlations observed of turnover magnitude $M^0$
and magnitude spread $\sigma$ with host galaxy luminosity
(e.g.\ \citealt{jordan2006}),
and also signs of peculiarities in the GCLFs of low surface brightness
dwarfs, mostly with excess low-mass clusters \citep{sharina2005,georgiev2009}.

The GCLFs of UDGs are relatively uncharacterized, with 
NGC~1052-DF2 and DF4 as well-known exceptions.
These two nearby galaxies with low to zero dark matter content
also have a remarkable excess of large, luminous star clusters with
$M_V \sim -9$ \citep{shen2021}.
In contrast, the nearby NGC~5846-UDG1 has a fairly normal GCLF 
(F606W-band $M^0 = -7.5$ and $\sigma = 1.1$; \citealt{danieli2022})\footnote{There were two different sets of GCLF values provided in the paper, but these are the correct ones (S.\ Danieli, priv.\ communication).}.
The stacked GCLFs of Virgo UDGs \citep{lim2020} 
and of Coma UDGs also appear fairly normal, with
F814W $M^0 \simeq -8.1$ and $\sigma \simeq 0.8$ for the latter
\citep{saifollahi2022}\footnote{The aperture correction in this study for F814W was 0.2~mag smaller than ours (see Section~\ref{sec:phot}), as they did not use an empirical PSF, 
which includes effects of the drizzling process and other variations.
Thus we interpret their result to be $M^0_\mathrm{F814W} \simeq -8.3$.}.

DGSAT~I differs from most of the UDGs above in having an excess of
$\simeq$~4--5 GCs brighter than expected from a canonical GCLF\footnote{DF07
may also have an excess of bright GCs,
and is the Coma UDG with a very narrow GC colour spread, like DGSAT~I.}.
In this respect it resembles DF2 and DF4, 
also in the unusually large sizes of its luminous, $\omega$~Cen-like clusters.
although perhaps
not for the same reason
(which in their case may again relate to a high-speed galactic collision;
\citealt{vandokkum2022a}).

The most promising formation scenario for DGSAT~I may be that of \citet{trujillo-gomez2021}, 
where early, intense star formation created a top-heavy cluster mass function.
This could be the same mechanism discussed above as potentially causing the narrow GC colour spread.
There is also a puzzle of why another well-studied UDG (NGC~5846-UDG1)
has a monochromatic, populous GC system and a suspected similar formation history to DGSAT~I, yet has a standard GCLF \citep{danieli2022}.

\subsection{Piecing together the clues}

The combined properties of DGSAT~I discussed above, and in the previous literature,
suggest that the galaxy is evolutionarily related to cluster UDGs.
Its GC system is relatively rich (when considering their total mass), 
is more compact than the stellar light, and
also has a red colour that matches the stellar light and
deviates from typical cluster dwarf trends.
These properties fit into the interpretation of some UDGs as failed galaxies
with overmassive dark matter haloes--which is supported by the high dynamical
mass of DGSAT~I as well as by its low metallicity and $\alpha$ enhancement.
This galaxy appears to represent a missing link for GC-rich
cluster UDGs that quenched in the field before infall
(e.g.\ \citealt{gannon2022}).
Even so, the unusually high luminosities and large sizes
of the GCs in DGSAT~I,
and their narrow spread of colours, point to diversity in the early SFHs even among the failed-galaxy sub-class of UDGs.

Other properties of DGSAT~I
are especially peculiar, and resist easy explanation.
The extreme chemical abundance pattern \citep{MN2019} remains
a mystery, as does the trace of recent star formation with no residual gas.
These quirks may actually represent additional important clues for understanding the nature and origins of this class of UDGs.
For example, Section~\ref{sec:backsplash} discusses some scenarios for the gas cycle and the recent star formation.

Finding additional analogues to DGSAT~I would also be essential in making progress
in our understanding.\footnote{The GC system of S82-DG-1, a red UDG in a void, was studied by \citet{roman2019}.  However, this galaxy is very likely to be a satellite of a massive spiral galaxy (NGC~1211) and is not actually an example of a UDG in a low-density environment.}
In particular, it is uncertain
if this galaxy is truly isolated or is a product of recent backsplash, 
while the environment may be clearer for other examples.
There were indeed two other quiescent UDGs in low-density environments discussed
in \citet{papastergis2017}: 
follow-up work has revealed that the distance was wrong for one of them
(R-127-1) and it is a nearby ``normal'' dwarf rather than a distant UDG,
while the other (M-161-1) remains a good analogue to DGSAT~I, with further analysis underway \citep{buzzo2022}.

\section{Summary and Conclusions}\label{sec:summ}

In this paper, we investigated the isolated, quiescent UDG DGSAT~I and its GC system using \textit{HST}/ACS imaging in the F606W and F814W filters, along with imaging from \textit{Spitzer}/IRAC. The results of this paper are as follows.

\begin{itemize}

\item The \textit{Spitzer}/IRAC photometry confirms that the blue overdensity has an optical--NIR colour unlike those of background galaxies. Likewise, we find the probability that it is a chance projection of a dIrr galaxy at a similar distance to be $\lesssim0.1\%$. 
The resulting conclusion that the overdensity is a region of younger (${\sim}500$ Myr) stars 
challenges the proposed backsplash nature of DGSAT~I, given the expected gas stripping.

\item We find the GC system of DGSAT~I to be more compact than the galaxy ($R_\mathrm{GC}/R_\mathrm{e} = 0.7$). This is an emerging trend in UDGs and opposite to what is found in other dwarf galaxies.

\item We measure a total number of GCs 
of $N_\mathrm{GC} \simeq 12 \pm 2$ and a total GC mass of $M_\mathrm{GC} \sim 6 \times 10^6~M_\odot$. Scaling relations between these parameters and halo mass
point to a halo that is normal or overmassive, respectively, for the galaxy's stellar mass.
The high velocity dispersion and dynamical mass of DGSAT~I \citep{MN2019} supports the overmassive halo, and we classify DGSAT~I as a ``failed galaxy''.

\item The GCLF of DGSAT~I is brighter, particularly in F814W, and much broader than in cluster dwarfs.
Both properties may be explained by a population of ${\sim}5$ overluminous and large GCs in DGSAT~I, leading to a top-heavy GCLF similar to those in NGC~1052-DF2 and DF4, though likely of different origin.

\item DGSAT~I has an identical colour to that of its GCs ($\mathrm{F606W} - \mathrm{F814W} = 0.46$, $V-I=1.03$). The redness of the GCs compared to GCs around other cluster dwarfs is another clue that DGSAT~I may reside in an overmassive halo.

\item The colour spread of the GCs around DGSAT~I is consistent with zero, suggesting that the GCs likely have identical ages and metallicities, and formed in a single intense starburst. This is consistent with the picture of \cite{trujillo-gomez2021}, where the formation of an extreme GC system is tied to the formation of the UDG, and may also explain the top-heavy GCLF.

\end{itemize}

Overall, a picture is emerging where the unique galaxy DGSAT~I may provide a window into the evolution of a sub-class of GC-rich cluster UDGs, where early starbursts and quenching created failed galaxies.

\section*{Acknowledgements}

We thank the anonymous referee for suggestions which improved the quality of the manuscript, and Shany Danieli, Bill Harris and Viraj Pandya for helpful discussions.
We thank NSERC for financial support.
SRJ acknowledges funding support from the Australian Research Council through Discovery Project grant DP200102574 during the course of this work.
AJR was supported as a Research Corporation for Science Advancement Cottrell Scholar.
DMD acknowledges financial support from the Talentia Senior Program
(through the incentive ASE-136) from Secretar\'\i a General de 
Universidades, Investigaci\'{o}n y Tecnolog\'\i a, de la Junta de
Andaluc\'\i a. DMD acknowledges funding from the State Agency for
Research of the Spanish MCIU through the ``Center of Excellence Severo
Ochoa'' award to the Instituto de Astrof{\'i}sica de Andaluc{\'i}a
(SEV-2017-0709) and project (PDI2020-114581GB-C21/ AEI /
10.13039/501100011033).
Based in part on observations made with the NASA/ESA \textit{Hubble Space Telescope},
obtained from the data archive at the Space Telescope Science Institute.
STScI is operated by the Association of Universities for Research in
Astronomy, Inc.\ under NASA contract NAS 5-26555.
Support for Program number HST-GO-14846 was provided through a grant from
the STScI under NASA contract NAS5-26555.
This research has made use of the NASA/IPAC Infrared Science Archive, which is
funded by the National Aeronautics and Space Administration and operated by
the California Institute of Technology.
This work is based in part on observations made with the {\it Spitzer Space Telescope},
which is operated by the Jet Propulsion Laboratory, California Institute of
Technology under a contract with NASA.
This research made use of Astropy,\footnote{\url{http://www.astropy.org}} a
community-developed core Python package for Astronomy \citep{astropy:2013,
astropy:2018}.
This research made use of Photutils, an Astropy package for detection and
photometry of astronomical sources \citep{bradley2021}.
This research has made use of the NASA/IPAC Extragalactic Database (NED),
which is operated by the Jet Propulsion Laboratory, California Institute of
Technology, under contract with the National Aeronautics and Space
Administration.
This research has made use of NASA’s Astrophysics Data System Bibliographic
Services.

\section*{Data Availability}

The \textit{HST} observations are available in MAST.
The \textit{Spitzer} images are available from the \textit{Spitzer} Heritage Archive.

\bibliographystyle{mnras}
\bibliography{ms} 

\bsp	
\label{lastpage}
\end{document}